\documentclass[Style/sn-mathphys]{Style/sn-jnl}
\theoremstyle{thmstyleone}%
\theoremstyle{thmstyletwo}%

\theoremstyle{thmstylethree}%

\graphicspath{{Figures/}}
\usepackage{amsmath}

\usepackage{etoolbox,xspace}

\newcommand{\definetextttcmd}[1]{\csdef{#1}{\texttt{#1}\xspace}}
\forcsvlist{\definetextttcmd}{PORTALS, CGYRO, TGLF, TGYRO, NEO, BoTorch, GPyTorch, ASTRA, TRANSP, GX, TANGO, GENE, TRINITY, QuaLiKiz, PyTorch, GKW, NCLASS, EPED, TORIC, FPPMOD}

\newcommand{\referee}[1]{{\color{black}#1}}

\begin{document}

\title[Performance in SPARC first campaign]{Core performance predictions in projected SPARC first-campaign plasmas with nonlinear CGYRO}

\author*[1]{\fnm{P.} \sur{Rodriguez-Fernandez}}\email{pablorf@mit.edu}

\author[1]{\fnm{N. T.} \sur{Howard}}

\author[1]{\fnm{A.} \sur{Saltzman}}

\author[1]{\fnm{L.} \sur{Shoji}}

\author[2]{\fnm{T.} \sur{Body}}

\author[2]{\fnm{D. J.} \sur{Battaglia}}

\author[1]{\fnm{J. W.} \sur{Hughes}}

\author[3]{\fnm{J.} \sur{Candy}}

\author[4]{\fnm{G. M.} \sur{Staebler}}

\author[2]{\fnm{A. J.} \sur{Creely}}

\affil[1]{\orgdiv{MIT Plasma Science and Fusion Center}, \city{Cambridge}, \state{MA}, \country{USA}}

\affil[2]{\orgdiv{Commonwealth Fusion Systems}, \city{Devens}, \state{MA}, \country{USA}}

\affil[3]{\orgdiv{General Atomics}, \city{San Diego}, \state{CA}, \country{USA}}

\affil[4]{\orgdiv{Oak Ridge National Laboratory}, \city{Oak Ridge}, \state{TN}, \country{USA}}

\abstract{
This work characterizes the core transport physics of SPARC early-campaign plasmas using the \PORTALS-\CGYRO framework \cite{rodriguez-fernandez_enhancing_2023}.
Empirical modeling of SPARC plasmas with L-mode confinement indicates an ample window of \textit{breakeven} ($Q>1$) without the need of H-mode operation.
Extensive modeling of multi-channel (electron energy, ion energy and electron particle) flux-matched conditions with the nonlinear \CGYRO code for turbulent transport coupled to the macroscopic plasma evolution using \PORTALS reveal that the maximum fusion performance to be attained will be highly dependent on the near-edge pressure.
Stiff core transport conditions are found, particularly when fusion gain approaches unity, and predicted density peaking is found to be in line with empirical databases of particle source-free H-modes.
Impurity optimization is identified as a potential avenue to increase fusion performance while enabling core-edge integration.
Extensive validation of the quasilinear \TGLF model builds confidence in reduced-model predictions.
The implications of projecting L-mode performance to high-performance and burning-plasma devices is discussed, together with the importance of predicting edge conditions.
}

\keywords{
PORTALS, SPARC, L-mode, predictions 
\\
\textit{Submitted for publication to: \textit{Phys. Plasmas}}
}

\maketitle

\section{Introduction}
\label{sec:Introduction}

Upcoming burning plasma experiments in both ITER \cite{Doyle2007} and SPARC \cite{Creely2020a,Rodriguez-Fernandez2022} are designed to achieve maximum fusion power and fusion gain ($Q$) in H-mode operation, with their baseline projections of $Q=10$ (ITER Baseline Scenario, IBS) and $Q=11$ (SPARC Primary Reference Discharge, PRD) necessitating a high energy confinement regime to sustain burning plasma temperatures at modest input power.

While operating plasmas in H-mode improves the energy confinement thanks to the formation of an edge transport barrier--- thus reducing the power requirements to sustain the confined plasma energy--- there are a number of unfortunate aspects that are associated with the formation of an edge \textit{pedestal}.
Operation in H-mode typically comes with associated edge localized modes (ELMs) that require either mitigation or suppression to avoid their detrimental effects on plasma facing components, which will be a pressing challenge for burning plasma experiments and fusion pilot plants  \cite{eich_elm_2017}.
Even though there are variations to H-mode operation that are ELM-free \cite{ViezzerE.2018}, the predictive capabilities to project access to those regimes and to predict confinement times in reactor-relevant conditions are very limited.
Similarly, I-mode regimes \cite{Whytea2010} (also characterized by an energy edge transport barrier, but without a particle transport barrier) that exhibit improved confinement relative to L-mode plasmas have limited predictive capabilities, as the underlying turbulence and transport mechanisms are subject of on-going research (e.g. \cite{bielajew_edge_2022, bielajew_edge_2023}).
Most predictive work ---either with the use of empirical confinement laws such as $\tau_E^{98y2}$ \cite{Doyle2007} or edge pedestal models such as \EPED \cite{Snyder2009}--- is focused on type-I ELMy H-modes, thus only providing an upper bound to performance predictions for future devices, assuming that operation without ELMs requires\referee{, typically,} a degradation of pedestal pressure from that coming from peeling-ballooning stability considerations \cite{Hughes2020a,Rodriguez-Fernandez2020a}.

While both H-mode and I-mode operation is associated with higher confinement levels and thus higher efficiency, every tokamak can operate in L-mode conditions.
L-mode can be described as the foundational and more naturally occurring state of plasma confinement in tokamak devices.
It is characterized by the lack of a significant edge transport barrier and has associated lower confinement levels. It is often assumed that the lower confinement in L-mode regimes is due to stronger turbulent transport, which is particularly important at the edge of the plasma, where shearing of turbulence is not strong enough (compared to H-mode regimes) to sustain a high cross-field gradient and transport barrier.

The upcoming SPARC tokamak will first operate with L-mode regimes, which are projected by empirical scalings to achieve \textit{breakeven} ($Q>1$) conditions. These high-performance L-modes are currently expected to be, arguably, the first time that $Q>1$ is attained in magnetic confinement fusion.
Because of this, the SPARC team is performing extensive investigations of several aspects of these L-mode plasmas \cite{battaglia_aps23,body_aps23,tinguely_aps23,rodriguezfernandez_aps23}, which focus on achieving optimal performance from the scenario. In this paper, we describe the progress so far in understanding the core turbulence characteristics of these high-performance, DT-fueled L-mode plasmas.
Leveraging the \PORTALS-\CGYRO framework \cite{rodriguez-fernandez_enhancing_2023}, here we present a comprehensive study of achievable core gradients and core performance in projected L-mode-like conditions in SPARC, and discuss the importance of edge conditions to attain fusion performance goals.

\section{Background}
\label{sec:Background}

SPARC \cite{Creely2020a,Rodriguez-Fernandez2022} is a high-field ($B=12.2T$), medium-size ($R=1.85m$), deuterium-tritium (DT) capable machine under construction by Commonwealth Fusion Systems in Devens, Massachusetts (USA), with construction to be completed by 2025 and first plasma in 2026. 
The accelerated timeline of SPARC is enabled thanks to novel high-temperature superconductor technology, which allows the operation at high magnetic field \referee{and} a moderate size.
The high magnetic field results in the capability to operate at high plasma current ($I_p\leq8.7MA$) and high absolute density ($\langle n_e\rangle\approx3\cdot10^{20}m^{-3}$ in the baseline H-mode regime)\referee{, while remaining} reasonably far from stability boundaries ($q_{95}\approx3.4$, $f_G=n_e/n_G\approx0.37$, $\beta_N\sim1.0$) \cite{Creely2020a,Rodriguez-Fernandez2020a,Rodriguez-Fernandez2022,Sweeney2020a}. These conditions are favourable for high energy confinement and fusion energy production, despite being smaller than a number of tokamaks that have been built and operated, such as JET \cite{mailloux_overview_2022}, TFTR \cite{french_construction_1983} and JT60-U \cite{kishimoto_advanced_2005}.
Projections of performance in SPARC indicate a maximum fusion gain in the range of $Q=8-11$ \cite{Creely2020a,Rodriguez-Fernandez2020a,rodriguez-fernandez_nonlinear_2022} operating with a mix of 50-50 DT fuel, provided that the pedestal remains at \EPED-predicted \cite{Hughes2020a} levels ---i.e. at ELMy H-mode levels--- with enough ELM control to ensure machine survivability.
Performance prediction in other regimes, such as L-mode, I-mode and other H-mode variations with somewhat degraded pedestal performance, is currently ongoing, although with less confident physics-based projections due to the lack of reliable predictive tools of edge and near-separatrix conditions.

Before targeting these high-performance H-mode discharges, however, SPARC will demonstrate \textit{breakeven} ($Q>1$) in a DT pulse as the conclusion of its first operational campaign. As well as demonstrating breakeven as soon as possible, the first campaign will need to minimize total neutron production so as to not preclude vessel entry for maintenance and upgrades. With these goals in mind, the SPARC team investigated several high-performance L-modes as potential breakeven candidate scenarios.

L-modes are desirable for several reasons. Unlike H-modes, L-modes do not have an input power threshold. Therefore ---if SPARC L-modes are found to have sufficient confinement--- $Q>1$ can be demonstrated with less fusion power and neutron production, while also reducing the number of ion cyclotron resonance heating (ICRH) antennas which must be commissioned to full-power in the first campaign. L-modes are also naturally ELM-free, which both avoids the need to commission the ELM mitigation system and reduces the risk of damage to plasma facing components (PFCs). L-modes are also more reproducible, have improved density control and less impurity accumulation than H-modes. However, L-modes also have significantly lower energy confinement than H-modes ---and so studying whether we can achieve sufficient performance to reach $Q>1$ in L-modes is crucially important.

Initial low-fidelity scoping of the SPARC L-mode operational space using a POPCON analysis\footnote{Available at \url{https://github.com/cfs-energy/cfspopcon}.} \cite{Houlberg1982} indicated that $Q>1$ can be achieved for a range of densities and heating powers, assuming an energy confinement given by the ITER-89P scaling \cite{Yushmanov1999}. To reduce the time required for scenario development, a family of candidate scenarios was developed for parameters which are relaxed slightly relative to the PRD. These scenarios are for single-null geometries\footnote{Available at \url{https://github.com/cfs-energy/SPARCPublic/tree/main/LmodeLowerSingleNull}.} at slightly reduced elongation (areal elongation $\kappa_A=1.7$ vs $\kappa_A=1.75$ in the PRD) and plasma current ($I_p=8.5MA$ vs $I_p=8.7MA$ in the PRD), for densities ranging from $\langle n_e\rangle\approx1.0\cdot10^{20}m^{-3}$ (near the predicted LOC-SOC transition \cite{Rice2020}) to $\langle n_e\rangle\approx2.1\cdot10^{20}m^{-3}$ \cite{battaglia_aps23}. These scenarios were projected by empirical scalings to achieve $Q>1$ while staying within operational limits such as those for disruption avoidance and PFC protection \cite{body_aps23}. This initial scoping was used to guide a series of higher-fidelity studies, investigating physics such as RF coupling and power exhaust. In this paper, we present high-fidelity core transport modelling for these high-performance L-modes --- investigating in unprecedented detail which of the candidate scenarios has the highest probability of achieving $Q>1$, as well as suggesting further optimization of the scenario to maximize performance.

\subsection{Physics-based modeling techniques}

To study core performance, this work leverages high-fidelity simulations using \PORTALS-\CGYRO and medium-fidelity simulations using \PORTALS-\TGLF.
\PORTALS \cite{rodriguez-fernandez_enhancing_2023} is a novel, open-source \cite{mitim} toolset developed to accelerate the numerical convergence of time-independent transport solvers.
The speedup achieved by the utilization of surrogate-based optimization techniques \cite{balandat_botorch_2020} in \PORTALS has enabled some of the highest fidelity predictions of core transport to date.
Direct nonlinear gyrokinetic simulations with the \CGYRO \cite{Candy2016} code have been used to predict steady-state (multi-channel flux-matched) plasmas in SPARC PRD \cite{rodriguez-fernandez_nonlinear_2022}, ITER IBS \cite{howard_prediction_2024}, DIII-D ITER Similar Shape \cite{howard_simultaneous_2024,rodriguez-fernandez_enhancing_2023}, ARC \cite{holland_aps23} and JET H, D and T Ohmic plasmas \cite{rodriguezfernandez_eps23,rodriguezfernandez_2024_jet}.

The quality of \PORTALS predictions depends on the model employed to simulate gradient-driven turbulent transport.
Drift-wave-type turbulence is the dominant core transport mechanism in the plasmas of interest, and is modeled here using the GPU-accelerated \CGYRO \cite{Candy2016} code.
\CGYRO is an initial-value, Eulerian, multi-species, spectral gyrokinetic solver designed specifically for electromagnetic, collisional, and multi-scale tokamak turbulence simulations. It solves the coupled gyrokinetic-Maxwell system of equations in the local $\delta f$ limit \cite{holland_examination_2021} and is optimized for scalability and performance on modern high-performance computing architectures \cite{candy_multiscale-optimized_2019}, including GPU-accelerated systems.
\CGYRO has been used to study a wide range of tokamak plasmas and is used in this work to predict the turbulent transport in flux-matched SPARC plasmas, to provide a high-fidelity prediction of the core temperatures and density.
On the other hand, the Trapped Gyro-Landau Fluid (\TGLF) quasilinear model of turbulent transport is used for the evaluation of performance at much reduced cost.
TGLF solves a linear system of gyro-fluid equations \cite{Staebler2005} and uses a saturation rule \cite{Staebler2007} to estimate the transport levels associated with the typical drift-wave-type micro-instabilities in the core of tokamaks.
\TGLF is benchmarked against \CGYRO predictions in this work to provide confidence in the reduced-model predictions.
In both the \PORTALS-\CGYRO and \PORTALS-\TGLF simulations, neoclassical transport is calculated with the \NEO code \cite{Belli2008}, but found to be negligible compared to turbulence as modeled by both \CGYRO and \TGLF ($<1.2\%$ in all flux-matched radii).

\section{Core turbulence modeling}

To produce physics-based predictions (in contrast with the empirically-based or data-driven predictions used for the initial scoping) of core performance, we start with the definition of a potential parameter space of operation of early-campaign SPARC plasmas.
With the goal of understanding the maximum fusion performance in SPARC L-modes, the scenarios presented in this section will be characterized by being operated with the full plasma current ($I_p=8.7MA$), magnetic field ($B_T=12.2T$) and shaping ($\kappa_{sep}=1.97$, $\delta_{sep}=0.54$). These parameters are assumed to be the same as for the PRD scenario, and assumed to provide an upper bound on expected performance, as plasma current, magnetic field and shaping parameters are expected to increase confinement time.
Similarly, impurity mix is assumed to be the same as for the PRD: fuel dilution $f_{DT}=0.85$, tungsten concentration $f_W=1.5\cdot10^{-5}$, He-3 minority concentration $f_{min}=0.05$ and effective ion charge $Z_\mathrm{eff}=1.5$ (for further information about these choices, the reader is referred to Refs~\cite{Creely2020a,Rodriguez-Fernandez2020a,Rodriguez-Fernandez2022}).
Auxiliary input power absorbed from ICRH is considered to be below $P_\mathrm{ICRH}<15MW$, consistent with commissioning plans for the machine and its actuator systems.
For conservatism and because of the lack of reliable predictive models, no core fueling is assumed and therefore particle flux-matching is equivalent to finding the null-flux condition, $\Gamma_e=0$.

\subsection{On parameter selection and assumptions}

In the parameter scans presented in following sections, three parameters are chosen as free variables and varied to shed light on the expected performance of SPARC early-campaign plasmas: total input power ($P_\mathrm{in}=P_\mathrm{ICRH}+P_\mathrm{OH}$), edge electron density ($n_\mathrm{edge}$) and edge temperature ($T_\mathrm{edge}$), the latter equal for ions and electrons.
In the context of core transport predictions, ``edge'' refers to the $\rho_\mathrm{tor}=0.9$ (square root of normalized toroidal flux) flux surface, which corresponds to $r/a\approx 0.943$ (normalized half-width of midplane intercept) and $\psi_n\approx0.922$ (normalized poloidal flux) for the plasmas of interest.
We must note that the only true controllable parameter from an operational point of view is the ICRH launched power. 
Operational density is somewhat controllable, but the ability to reach desirable edge density levels will depend on the balance of fueling and transport near the separatrix and the scrape-off layer, which remains an open question and out of the scope of this work focused on core transport predictions.
Similarly, edge temperature is a highly uncertain parameter that will be determined by the balance of transport, radiation losses and energy flows in the near-separatrix region. Due to the difficulties predicting the edge temperature, the approach taken here is to assume a potential range of edge temperatures (based on past experiments, as will be described in Section~\ref{sec:BC}) and explore the sensitivity of fusion gain and fusion power on the achievable edge pressure.

In summary, the parameter space is defined by the following ranges: $T_\mathrm{edge}=0.9-2.3keV$, $n_\mathrm{edge}=0.77-2.05\cdot10^{20}m^{-3}$ and $P_\mathrm{in}=8.3-18.3MW$. There will be an exploratory, very high density case ($n_\mathrm{edge}=4.0\cdot10^{20}m^{-3}$) that was run at low $T_\mathrm{edge}$ and high $P_\mathrm{in}$. To make this case survive, the tungsten concentration was halved, otherwise the plasma would radiatively collapse with the nominal tungsten content.
These ranges can be visualized in Figure~\ref{Fig:database}a) and b) (to be discussed in following sections), where the 12 different combinations of $(n_\mathrm{edge},T_\mathrm{edge},P_\mathrm{in})$ are shown.

During the parameter scans, the plasma geometry will be assumed to be fixed, including the internal flux surfaces. This assumption is motivated by: 1) the prohibitive computational cost of performing predictions of scenarios with different geometry (as the \PORTALS surrogate re-utilization technique would not be applicable), and 2) the expected small effect of internal equilibrium on transport, given the low normalized pressure, $\beta_N\lesssim0.5$, and expected small differences in Shafranov shift for all cases studied.
The power deposition is assumed to remain unchanged as well, both  the spatial location of power delivered to ions and electrons, and the ICRH power partition between species.

\begin{figure}
	\centering
	\includegraphics[width=1.0\columnwidth]{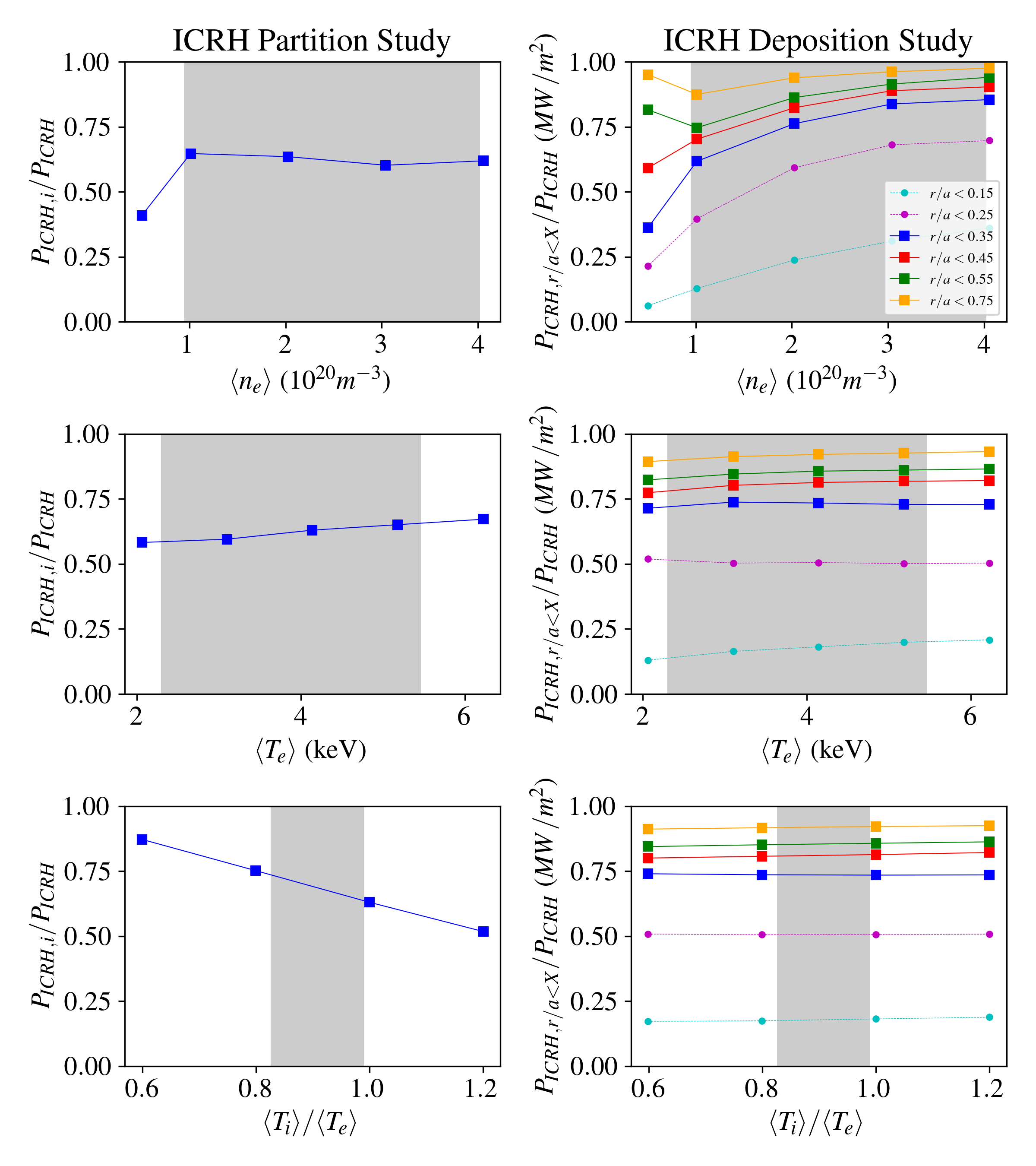}
	\caption{\TRANSP interpretive analysis of ICRH power delivery to plasma. Left column depicts the change in the fraction of ICRH power delivered to ions as a function of density, temperature and temperature ratio. Right column shows the change in the ratio of power delivered within the $r/a=[0.15,0.25,0.35,0.45,0.55,0.75]$ flux surfaces as a function of the same parameters.
    Shaded regions indicate the range of variation of the parameters in the predictive results from this work.
	When not varied, $\langle n_e\rangle=1.5\cdot10^{20}m^{-3}$, $\langle T_e\rangle=4.0keV$ and $T_i/T_e=1.0$ are assumed.
    }
	\label{Fig:transp}
\end{figure}

To study the sensitivity of the ICRH power deposition so that we can make the parametric scans easier while retaining any relevant physics, scans using interpretive \TRANSP \cite{Breslau2018} with various potential kinetic profiles were performed.
\TRANSP is run with the \TORIC full-wave code \cite{Brambilla1999} coupled with the bounce-averaged Fokker-Planck \FPPMOD model \cite{Hammett1986} to calculate the self-consistent response of the minority species to the wave field.
Figure~\ref{Fig:transp} shows the results of \TRANSP scans with various volume-averaged densities, volume-averaged temperatures, and temperature ratios.
All the rest of parameters are kept fixed and are similar to the \TRANSP simulations presented for the PRD \cite{Rodriguez-Fernandez2020a}, although interpretively.
The results show that the relative amount of ICRH power absorbed by the ions remains within the range $P_{\mathrm{ICRH},i}/P_\mathrm{ICRH}\sim0.65-0.75$ for the cases studied here, and therefore we considered it is a safe assumption to keep the power deposition fixed.
Similarly, the change in spatial deposition is not significant, with the strongest effect at the inner most location simulated, $r/a<0.35$ (motivated by the guidelines presented in Ref~\cite{rodriguez-fernandez_enhancing_2023}).
While the change in power deposition at that location is not completely negligible ($\sim25\%$ increase in power when doubling the density from $\langle n_e\rangle=1.0\cdot10^{20}m^{-3}$ to $2.0\cdot10^{20}m^{-3}$), the very stiff transport expected in the core will result in a very small change in the core temperature with variations in transported power.

We must note that the changes in both the ICRH partition and the deposition location are dominated by the changes in energy exchange power from the minority species, not the actual ICRH direct absorption by the minority or the plasma species.
Because the temperature of the background minority species (He-3) is assumed to be equal to that of the background ions when no ICRH power is applied, and because both the electron and ion temperatures are fixed throughout the interpretive \TRANSP simulations, there is artificial channeling of ICRH power from/to electrons to/from ions when the difference between $T_e$ and $T_i$ is too high for a given density choice.
This suggests that an important future update to the \PORTALS framework will be to allow for the auxiliary power delivery to evolve during the convergence process, as the temperature profiles are updated.

While the previous analysis has focused on the ICRH power, the plasmas of interest will also have a significant fraction of Ohmic heating.
This work assumes fixed Ohmic power (with a representative database value of $3.3MW$), despite its potential to change with density, temperature and impurity content.
It is because of this that the power scans are, in actuality, performed with respect to the total absorbed power, $P_\mathrm{in}$, which will be varied between $8.3MW$ ($5MW$ ICRH + $3.3MW$ of Ohmic) and $18.3MW$ ($15MW$ ICRH + $3.3MW$ of Ohmic).
The limitation of running with fixed Ohmic power is a consequence of the capabilities of the current \PORTALS framework, and will be improved in future work.

\subsection{On the choice of boundary condition}
\label{sec:BC}

As described earlier, both the edge density and edge temperature are assumed, imposed parameters in the core modeling.
The edge density assumption is such that the operating volume-averaged Greenwald fraction is scanned in the range $f_G=0.11-0.26$. This work assumes that fueling systems will be capable of sustaining such edge density\footnote{We must note that the edge-separatrix region, $r/a=0.943-1.0$, has no effect on core performance from the perspective of flux-tube local gyrokinetic turbulence and the assumptions employed in this work.}.

The determination of edge temperature and its validity is more subtle.
Transport predictions from the separatrix in L-modes, while possible \cite{angioni_dependence_2023,Angioni2022}, are more complex, uncertain and have a somewhat weaker physics basis, as L-mode edge transport remains an active area of research and experimentation and the validity of current transport models is still under investigation.
Radiation and impurity effects, the dynamics of neutrals, the effect of rotation and sheared flows, charge exchange processes, and the influence of equilibrium geometry (strongly shaped plasmas in double-null) near the separatrix introduce complexities that are difficult to predict for breakeven plasma conditions.
Furthermore, in addition to these, performance predictions that span the entire confined plasma must be coupled to scrape-off layer and divertor simulations and the difficulty to perform nonlinear gyrokinetic simulations at the edge (large normalized logarithmic gradients lead to small simulation time-steps and these locations require higher radial resolution) drove the decision of using $r/a=0.943$ as the boundary condition for the kinetic profiles, as a tunable parameter to explore the operational space.

\begin{figure}
	\centering
	\includegraphics[width=0.7\columnwidth]{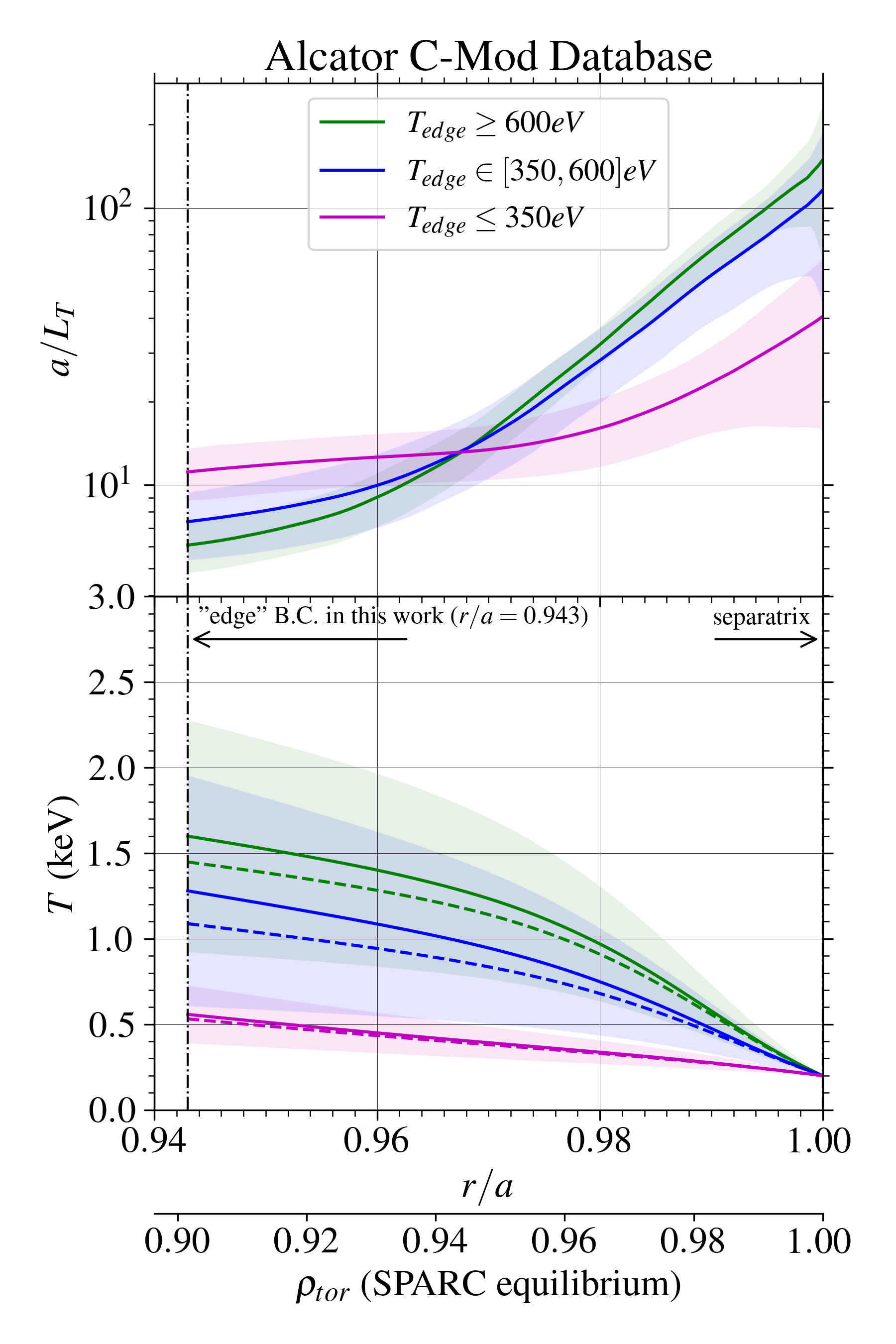}
	\caption{
 (top) Distribution (mean and standard deviation) of normalized logarithmic gradient in database of Alcator C-Mod non-H-mode plasmas in the edge region, $r/a=0.94-1.0$. Distributions are separated by the value of edge temperature.
 (bottom) Distribution of edge temperatures when the gradients are radially integrated from a boundary condition of $200eV$ at $r/a=1.0$ (dashed is the integration of the mean, solid is the average of the integration of upper and lower bounds.)
    }
	\label{Fig:bc}
\end{figure}

We chose $T_\mathrm{edge}=1.6\pm0.7$ keV (as a reminder, edge in this work refers to $r/a=0.943$) as the nominal assumption, motivated by a database study of non-H-mode plasmas in Alcator C-Mod.
A set of $\sim$2000 discharges in Alcator C-Mod were chosen by their stationarity. Time windows of at least 250 ms were extracted from stored Alcator C-Mod data where the variation in the magnetic field, plasma current, and line-averaged density was less than 5$\%$ during the time window. If ICRF power was present, it was required to vary no more than 7.5\% during the selected time window. These requirements ensured relatively steady discharge conditions while not being overly restrictive.
From the set of discharges satisfying these criteria, only plasmas with $q_{95}<4.5$ and without significant edge density gradient were considered so as to only explore non-H-mode-like kinetic profiles (relatively few discharges with high density gradient and $q_{95}<4.5$ met the above stationarity conditions in Alcator C-Mod). Outliers (plasmas with unphysically high temperature gradients due to bad fitting or bad experimental data) were removed.
As a result, a total of $\sim$600 discharges were used to derive distributions of achievable $T_{e,edge}$ (as measured by Thomson Scattering \cite{hughes_thomson_2003}) and associated $a/L_{T_e}$ profiles.
From this database, plasmas with $T_{e,edge}>600$ eV were chosen to be ``high-performance'' non-H-mode plasmas in Alcator C-Mod (this down-selection resulted in 61 discharges).
The distribution (mean and standard deviation) of $a/L_{T_e}$ in the edge region, $r/a=0.94-1.0$, is shown in the top panel of Figure~\ref{Fig:bc}.

The bottom panel of Figure~\ref{Fig:bc} shows the distribution of edge temperatures when gradients are radially integrated from an assumed separatrix temperature of $200eV$ at $r/a=1.0$ (derived from a two-point model), which was chosen as a potential operational boundary condition for the SPARC plasmas of interest.
This workflow therefore suggested ---under the assumption of a constant normalized gradient scale length, which implies similarity in the transport physics in that region between Alcator C-Mod and SPARC--- that $T_\mathrm{edge}=1.6\pm0.7$ keV (from the integration of ``highest``-performing plasmas in the Alcator C-Mod database, $T_\mathrm{edge}>600$ eV) is a reasonable choice for the edge temperature in this exploratory study. The validity of this edge assumption will be the subject of ongoing studies but the approach in this work was to attempt to identify a plausible set of L/I-mode edge profiles that were derived from existing experiment. We note however, that recent work by \textit{Hatch et al.} suggests that the edge region of I-mode is consistent with an electron-temperature-gradient mode limited edge. Further investigation into these findings is the subject of ongoing work, and could provide a more physically robust manner for predicting near-edge profiles in SPARC.

We acknowledge the various caveats that this workflow introduces.
First, the choice of normalized gradients scale lengths found in present-day tokamaks as a reference for SPARC could be subject to criticism, as it may as well be that SPARC (e.g. with higher plasma current and magnetic field among other parameters) could achieve higher gradients and therefore edge performance than those found in, e.g., Alcator C-Mod.
Secondly, the database used to derive the edge temperature does not differentiate between L- or I-mode regimes. Power threshold and regime access considerations are not being taken into account.
Finally, the choice of a separatrix temperature of $200eV$ (and equal to electrons and ions) is subject to variations depending on the operational regime (particularly the density and input power choice) and could be prone to optimization if it is indeed found to be a critical parameter to determine overall plasma performance.
While all these are valid limitations for these predictions, the work presented here will discuss the implications of these choices and the potential avenues for improvement in future work (particularly as discussed in Section~\ref{sec:Discussion}).

\subsection{High-fidelity modeling setup}

The \PORTALS technique for steady-state modeling presents an important advantage (apart from higher nominal computational efficiency) to standard methods: the possibility to re-utilize surrogates when predicting scenarios with variations in certain input parameters.
As explained in detail in Ref.~\cite{rodriguez-fernandez_enhancing_2023}, parameters that only affect the macroscopic evolution but not the local gradient-driven turbulence simulations (such as input power) and parameters whose effect is already included in nominal surrogates (such as density and temperature boundary conditions affecting core collisionality, temperature ratios and normalized pressure) are suitable for surrogate re-utilization.
To this end, we exploited this capability to study 12 scenarios that required only a total of 71 profile evaluations to achieve flux-matched conditions in electron energy, ion energy and electron particle fluxes, thus resulting in the prediction of electron temperature, ion temperature and electron density.
\referee{\PORTALS is currently implemented as an electron-solver, thus evolving the electron density and adjusting the densities of the different ion species by keeping a constant, radially uniform concentration.
This limits the capability of the framework to account for non-trace, multi-species particle transport (such as the effect of ash accumulation or radial variations of the D-T fraction).
Future developments of \PORTALS will include the capability to evolve the ion densities self-consistently.}

Six radial locations are chosen to describe the logarithmic gradient profiles, $r/a=[0.35,0.55,0.75,0.875,0.9,0.943]$. This choice is motivated by the database analysis presented in Ref.~\cite{rodriguez-fernandez_enhancing_2023} that suggested that 5 locations are enough to study inductive, on-axis heated discharges, \referee{and that the inclusion of additional radial locations between $r/a=0.35$ and the magnetic axis results in minimal changes to the total fusion power in SPARC.}
The additional location at $r/a=0.943$ was added to ensure proper description of L-mode-like gradient scale length profiles, which are expected to have more structure than H-mode-like profiles.
During the process to achieve steady-state conditions, target fluxes were updated to account for the change in radiation losses (Bremsstrahlung, line and synchrotron radiation), collisional energy exchange between electrons and ions and alpha power to electrons and ions.
The models employed for these target fluxes are identical to those in Ref.~\cite{rodriguez-fernandez_nonlinear_2022}.

With the total of 6 radial locations per profile and with the goal of achieving convergence in the three described transport channels, this comprehensive study required 426 (71 profiles $\times$ 6 local radii) flux-tube nonlinear gyrokinetic simulations, which were performed with the \CGYRO \cite{Candy2016} code.
Advances in GPU acceleration of \CGYRO allowed us to perform this study with a total computational cost of $\sim$80,000 GPU hours ($\sim$20,000 node hours), performed on the GPU partition of the NERSC Perlmutter supercomputer, equipped with NVIDIA A100 (40GB) GPUs.

All gyrokinetic simulations described in this paper were performed with the \CGYRO code \cite{Candy2016}. \CGYRO is a local, Eulerian gyrokinetic code that is optimized for multi-scale and near-edge simulation of tokamak discharges. The resolutions used in this work were relatively standard for the simulation of low-k (ion-scale) turbulence typically found in the core of tokamaks ---such as micro-tearing (MTM), ion temperature gradient (ITG) and trapped electron mode (TEM) turbulence. High-k TEM and electron temperature gradient (ETG) mode contributions were ignored in this work motivated in part by the large (greater than 1) target values of $Q_i/Q_e$ and the so-called fingerprint argument presented by \textit{Kotschereuther et al.} \cite{Kotschenreuther2019}. 
All simulations utilized relatively high fidelity physics including geometry represented by the Miller Extended Harmonic \referee{(MXH)} formulation \cite{Arbon2020}, fully electromagnetic turbulence ($\delta \phi$, $\delta A_{\parallel}$, and $\delta B_{\parallel}$), Sugama collisions \cite{sugama_nonlinear_1998}, and 4 gyrokinetic species: D, T, lumped impurity (Z = 4, A=8), and electrons. The lumped impurity species is meant to represent an average impurity that includes contributions from He-3, W, and other low-Z species. 
Rotation was not included in these simulations as no reliable estimate of the profile was available. 
Simulation domains varied slightly over the simulated radial regions ($r/a = 0.35, 0.55, 0.75, 0.875, 0.9, 0.943$) but nominal box sizes were $[L_x,L_y] \sim 125 \times 120\rho_s$ utilizing 24 \referee{complex} toroidal modes (\referee{$N_\alpha$}) and 512 radial modes (\referee{$N_r$}) across the domain to represent turbulence from $k_\theta \rho_s=0.0$ to $1.22$. Simulations also used 8 energy grid points (\referee{$N_{u}$}), 24 pitch angles (\referee{$N_\xi$}), and 24 points in theta (\referee{$N_\theta$}). Slightly higher radial and energy resolutions of approximately $\referee{N_r} = 630$ and $\referee{N_{u}}=12$ were used at $r/a =0.9$ and $0.942$. 

We must emphasize that the results presented here are not results of surrogate modeling of gyrokinetic turbulence, but of \textit{direct} nonlinear \CGYRO modeling.
Surrogates are just used to accelerate multi-channel ($Q_e$, $Q_i$, $\Gamma_e$) convergence, but the steady-state predictions ($T_e$, $T_i$, $n_e$) are a full-physics result (of course, under the constraints of ion-scale nonlinear gyrokinetics as embedded in \CGYRO) and must be interpreted as such.

\begin{figure}
	\centering
	\includegraphics[width=1.0\columnwidth]{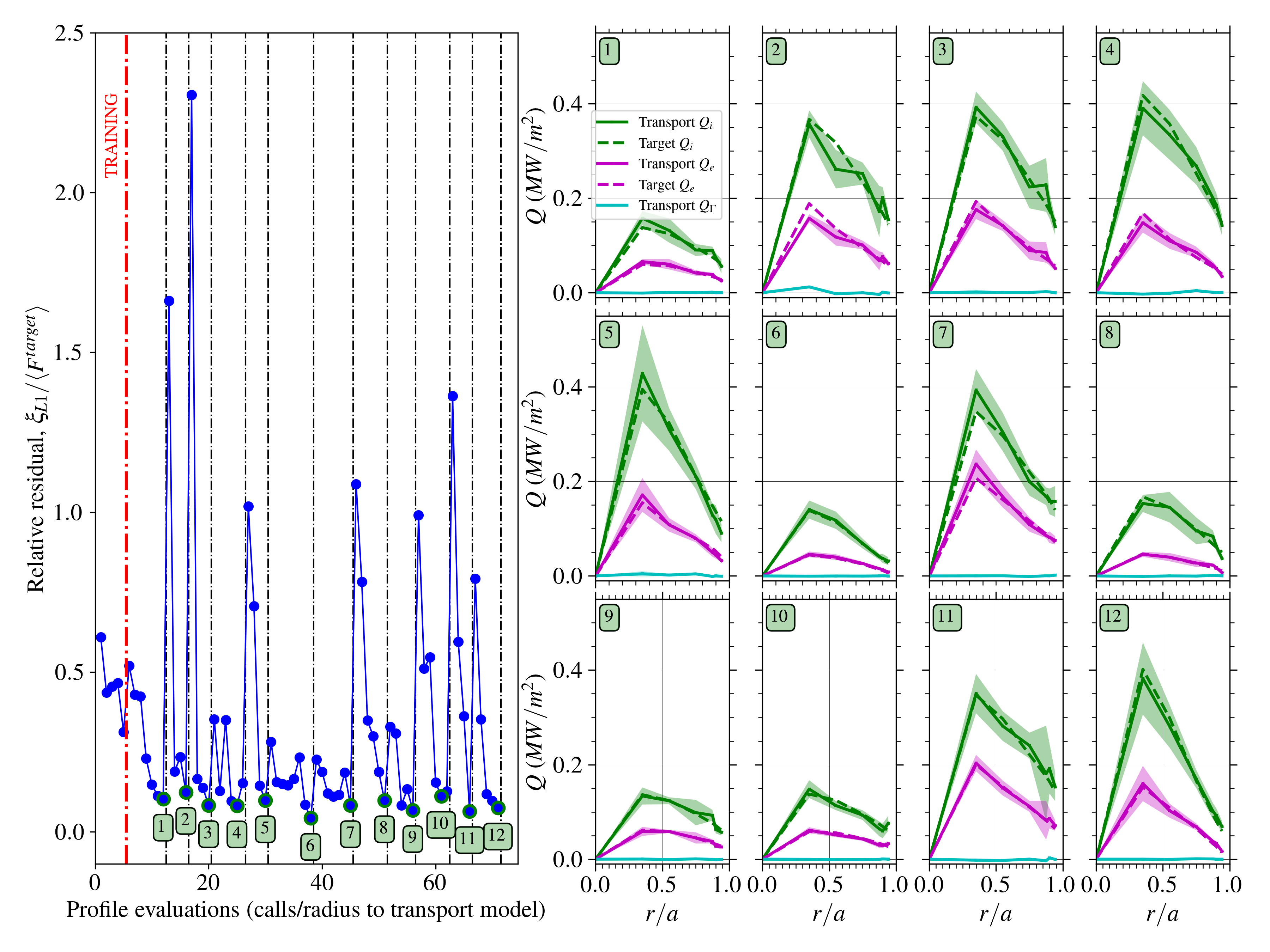}
	\caption{Summary of \PORTALS-\CGYRO multi-channel convergence.
    (left) $L_1$ residual normalized to the radial-average of the target flux per iteration, with the 12 converged cases numbered in green.
    (right) Transport and target flux profiles for the 12 converged cases, including the three channels: ion energy flux, electron energy flux and electron particle flux (transformed into energy units as a convective flux, $Q_{\Gamma}=\frac{3}{2}T_e\Gamma$). Shaded area indicates $\pm2\sigma$ as estimated for the \CGYRO transport fluxes.}
	\label{Fig:restarting}
\end{figure}

Figure~\ref{Fig:restarting} shows an overview of the \PORTALS process for the prediction of the 12 core scenarios.
\referee{
On the left, the multi-channel, multi-radius residual (i.e. the $L_1$, scalarized residual that accounts for electron energy, ion energy and electron particle flux residuals at all radii simulated) is plotted vs. the iteration number.
The 12 converged cases are numbered in green, with vertical dashed lines indicating the transition from one scenario to another, reutilizing flux surrogates.
On the right, the transport and target flux profiles (including the three channels) for the case with the lowest residual per scenario simulated are shown.
}
Convergence is achieved for each case when the residual is sufficiently low, and confirmed by visual inspection and expert judgement. 
\referee{Details on the residual definitions and on the convergence metrics are described in Ref.~\cite{rodriguez-fernandez_enhancing_2023}}.
Future work with \PORTALS will focus on automation of convergence metrics and more robust convergence criteria. However, considering the stiff and critical gradient behavior of the plasmas studied here, exact flux-matching (e.g. in case \#2) is not required for performance predictions, as extensively discussed in Refs.~\cite{rodriguez-fernandez_nonlinear_2022,rodriguez-fernandez_enhancing_2023}.

\section{Nonlinear gyrokinetic results}
\label{sec:PORTALScgyro}

\begin{figure}
	\centering
	\includegraphics[width=1.0\columnwidth]{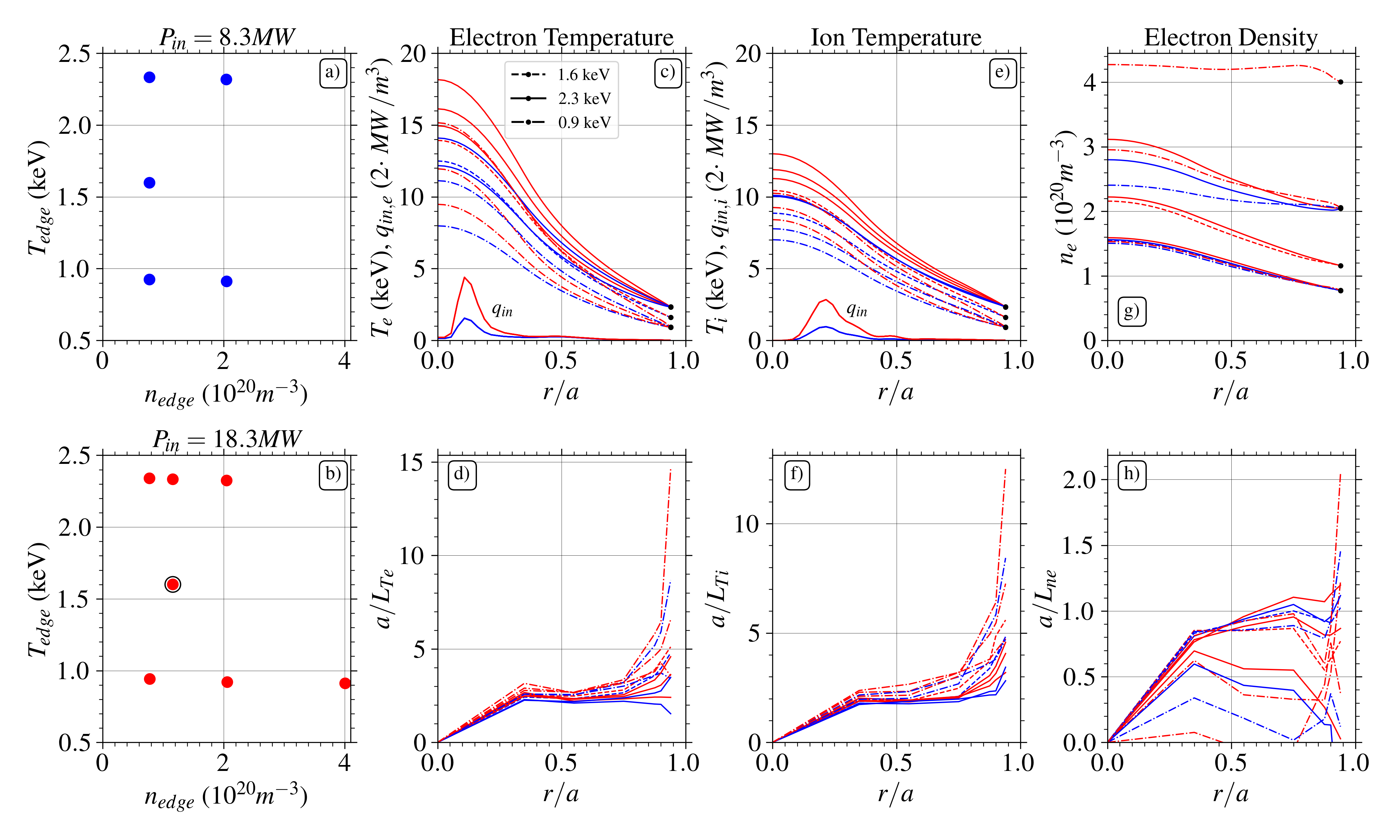}
	\caption{Summary of database of \PORTALS-\CGYRO profile predictions. 
    a) and b) depict the 12 scenarios chosen from combinations of the three input parameters (input power, edge temperature and edge density).
    c), e) and g) show the electron temperature, ion temperature and electron density prediction from the $r/a=0.943$ boundary condition.
    c) and e) also show the deposited power density for electrons and ions respectively (only two cases studied, $P_\mathrm{in}=8.3MW$ and $P_\mathrm{in}=18.3MW$, with fixed shape and species partition).
    d), f) and h) show normalized logarithmic gradients of the profiles. 
    In all figures, red and blue are used to indicate low input power ($P_\mathrm{in}=8.3MW$) and high input power ($P_\mathrm{in}=18.3MW$) respectively\referee{, and different line styles for the temperature boundary conditions}, for better visualization.
    In b), point with $P_\mathrm{in}=18.3MW$, $n_\mathrm{edge}=1.2\cdot10^{20} m^{-3}$ and $T_\mathrm{edge}=1.6 keV$ is highlighted with a black circle, as it is the scenario chosen for studies of the impurity effect in Section~\ref{sec:impurity}.
    }
	\label{Fig:database}
\end{figure}

Figure~\ref{Fig:database} summarizes the predictions of the three kinetic profiles.
Electron temperature is found to always be higher than the ion temperature\footnote{In this work, the only assumption that could somewhat change the temperature ratio is the assumed $T_i/T_e$ at $r/a=0.943$, which has only been given the value of $1.0$. Future work could explore sensitivity to this assumption, but we do not expect a strong deviation from $1.0$ given the high edge collisionality.}.
Gradient profiles show evidence of stiff inner core ($r/a<0.7$) and non-stiff edge conditions ($r/a=0.7-0.943$), which is reminiscent of experimental observations of L-mode plasmas \cite{Sauter2014}.
The case with very high density, that resulted in a volume-averaged Greenwald fraction of $f_G\approx0.47$ required a negative electron density gradient to achieve the null-flux condition in the plasma core, hence the slightly hollow electron density profile at mid-radius.

\begin{figure}
	\centering
	\includegraphics[width=1.0\columnwidth]{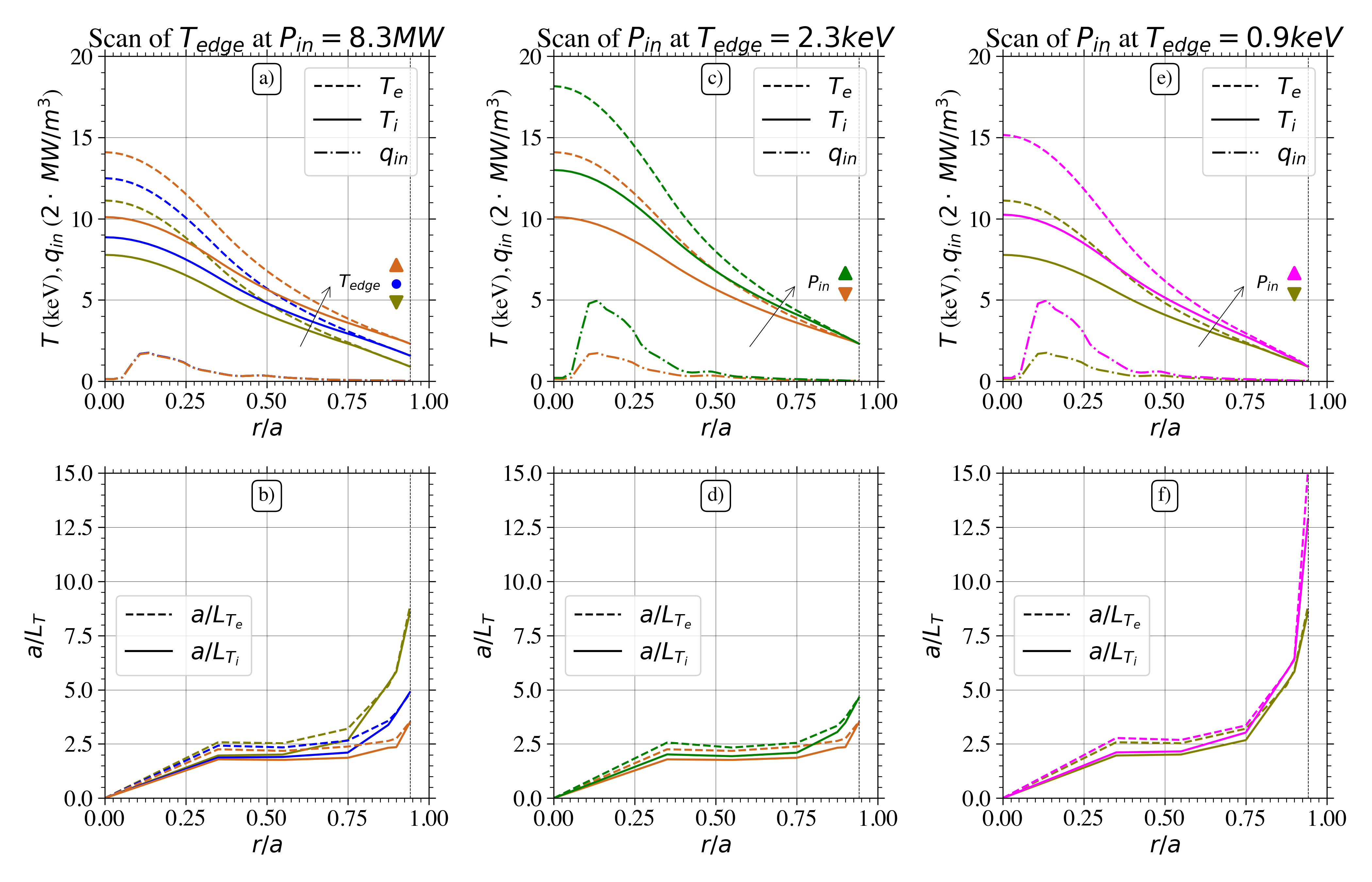}
	\caption{Exploration of core transport dynamics (predicted electron and ion temperatures together with power deposition profile at the top and normalized logarithmic gradients at the bottom) at fixed edge density, $n_\mathrm{edge}=0.77\cdot10^{20}m^{-3}$.
    a,b) Scan of edge temperature at low input power ($P_\mathrm{in}=8.3MW$), c,d) scan of input power at high edge temperature ($T_\mathrm{edge}=2.3keV$), and e,f) scan of input power at low edge temperature ($T_\mathrm{edge}=0.9keV$).
    }
	\label{Fig:stiff}
\end{figure}

\begin{figure}
	\centering
	\includegraphics[width=1.0\columnwidth]{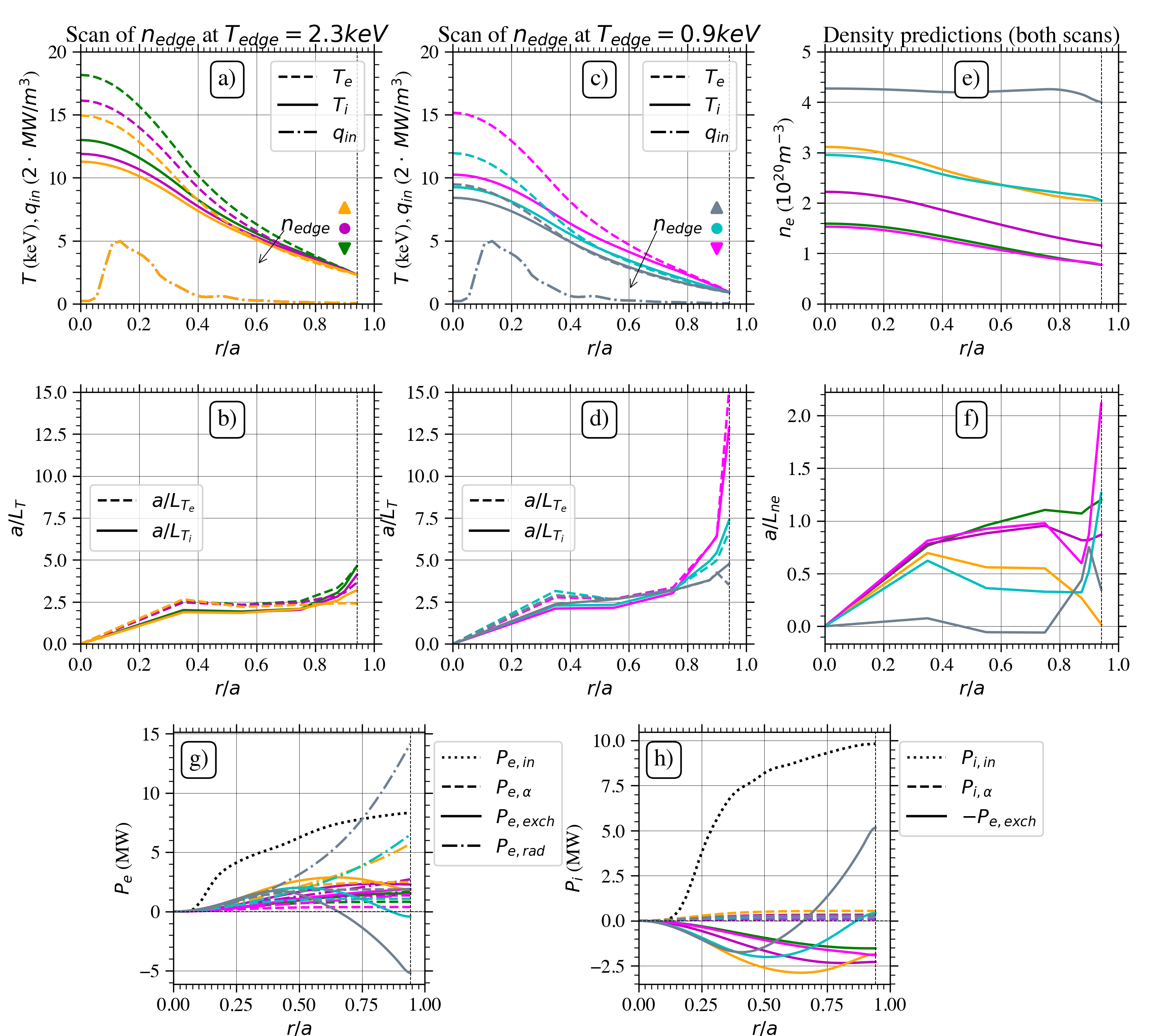}
	\caption{Scan of edge density at a,b) high ($2.3keV$) and c,d) low ($0.9keV$) edge temperatures, both at high input power, $P_\mathrm{in}=18.3MW$.
    Temperature profiles are shown at the top and logarithmic gradients at the bottom.
    e,f) compilation of density profiles and their gradients for the two scans.
	\referee{
	g,h) electron and ion sources (input power, alpha power), sinks (radiation losses) and exchange power (electron to ion energy exchange).
	}
    }
	\label{Fig:stiffdens}
\end{figure}

The observation of core transport stiffness from Figure~\ref{Fig:database} is further explored in Figures~\ref{Fig:stiff} and \ref{Fig:stiffdens}.
Figures~\ref{Fig:stiff}a) and b) show the exploration of effect of the edge temperature at low input power ($P_\mathrm{in}=8.3MW$) and low edge density ($n_\mathrm{edge}=0.77\cdot10^{20}m^{-3}$), indicating a clear flattening of the logarithmic gradient as the boundary condition is increased.
At high edge temperatures, the plasma becomes stiffer, mostly as a consequence of the increased gyro-Bohm unit, $Q_{GB}\propto T_e^{5/2}$.
For similar power (\referee{$Q$} in $MW$) to be exhausted via turbulent transport, the lower gyro-Bohm normalized energy flux \referee{(lower $Q/Q_{GB}$ when $Q_{GB}$ is higher)} requires a lower logarithmic gradient, which flattens the gradient profile.
This brings up an interesting corollary: under the assumption of local, flux-tube gyrokinetic turbulence, the parameter that may separate L-mode from H-mode is the absolute temperatures and densities, which enter through the gyro-Bohm normalization in transport solvers.
This means that, high-temperature L-modes ---as those expected in high-confinement, burning-plasma machines--- may behave more similarly to present-day H-modes. 
However, as it will expanded throughout this section, for the energy confinement time to be at levels predicted by L-mode scaling laws in high-performance plasmas, we require a high edge pressure, which may mean the formation of an edge transport barrier might be required.

Figures~\ref{Fig:stiff}c) and d), and  Figures~\ref{Fig:stiff}e) and f) show the exploration of the effect of the input power ($8.3MW$ and $18.3MW$) at high ($T_\mathrm{edge}=2.3keV$) and low ($T_\mathrm{edge}=0.9keV$) edge temperatures respectively, both at low edge density ($n_\mathrm{edge}=0.77\cdot10^{20}m^{-3}$).
At high edge temperatures, the plasma is stiffer, and the edge logarithmic gradient is not very sensitive to the input power.
In fact, in these stiff conditions, the higher input power ($18.3MW$ v.s. $8.3MW$) resulted in a $\sim15\%$ lower fusion gain because the fusion power did not increase by the equivalent amount (only $\sim90\%$).
On the contrary, at low edge temperatures, the increase in input power leads to a $\sim5\%$ increase in fusion gain.
We must emphasize, however, that these results assume that the same edge temperature is achieved at the two different power levels.

Similarly, Figures~\ref{Fig:stiffdens}a) and b) demonstrate the almost invariability of the temperature profile when density is varied at high edge temperature ($T_\mathrm{edge}=2.3keV$).
At low edge temperature ($T_\mathrm{edge}=0.9keV$), however, the temperature profile is more sensitive to density, with the plasma becoming colder as density is increased, as shown in Figures~\ref{Fig:stiffdens}c) and d).
In both cases the fusion gain is strongly affected by the density, increasing by a factor of $\sim3$ from $n_\mathrm{edge}=0.77\cdot10^{20}m^{-3}$ to $n_\mathrm{edge}=2.05\cdot10^{20}m^{-3}$.
We must emphasize here as well that these results assume that the same edge temperature is achieved at the same input power regardless of the edge density, thus assuming a higher edge pressure at higher density for the same input power.
Figures~\ref{Fig:stiffdens}e) and f) show the density profiles and their gradients predicted in both scans.
\referee{During these density scans, there is significant variation in the target fluxes per species, with the energy exchange power between electrons and ions and the radiation losses being the most sensitive to the density variation, as shown in Figures~\ref{Fig:stiffdens}g) and h).
With the assumption of constant impurity concentration, the radiation losses increase with density, consistent with the model implemented in \PORTALS and described in detail in Ref.~\cite{rodriguez-fernandez_nonlinear_2022}.
The trend of the electron to ion energy exchange power is more complex, with a non-monotonic radial variation and including a change of direction (when $T_i>T_e$ in the outer part of the simulated plasma region) at the highest densities.
}

\begin{figure}
	\centering
	\includegraphics[width=1.0\columnwidth]{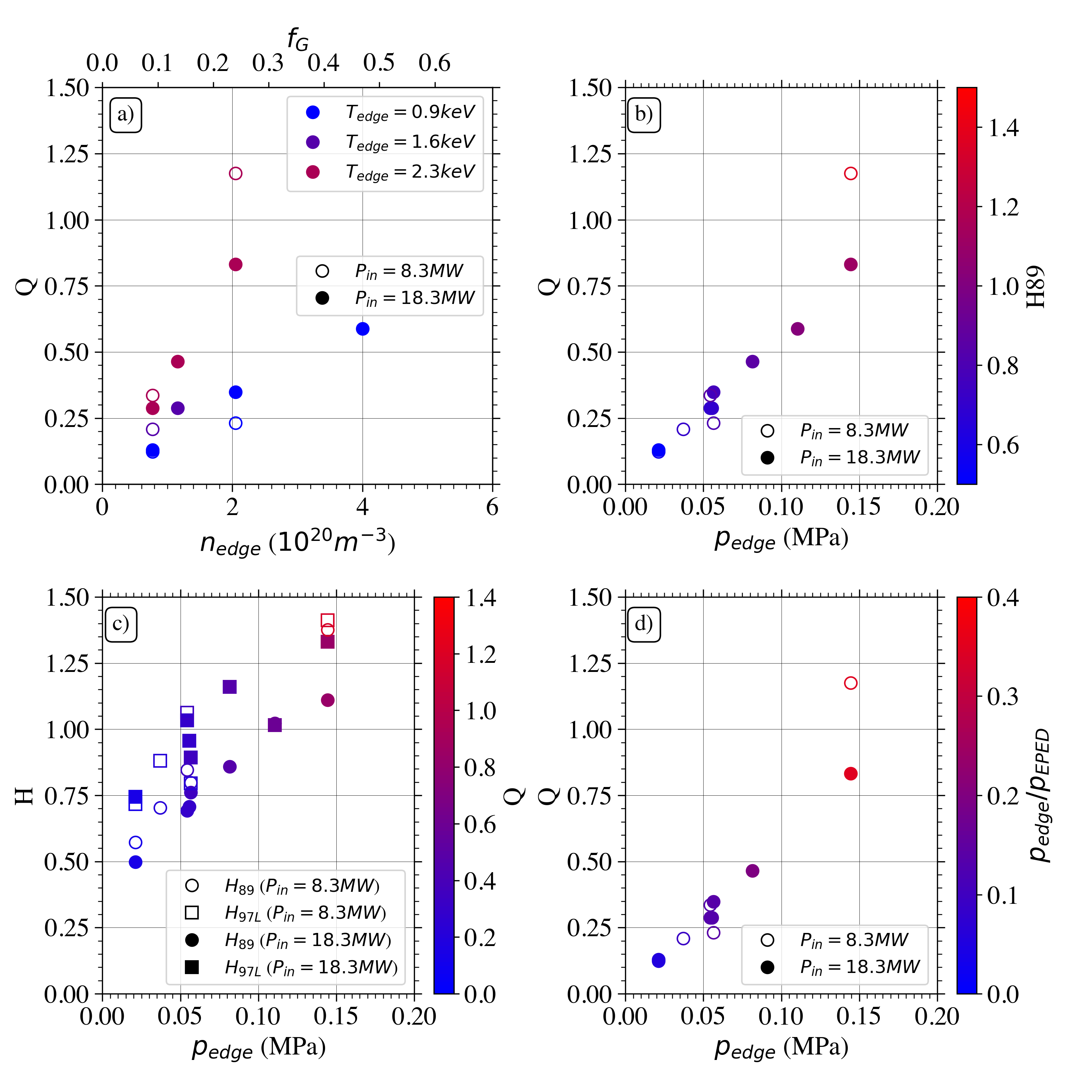}
	\caption{Study of fusion gain for the 12 scenarios considered.
	a) Fusion gain as a function of edge density, edge temperature and input power as separate free parameters.
	b) Fusion gain as a function of edge pressure and input power, with colorbar indicating the level of energy confinement over the $\tau_E^{89p}$ confinement law (H factor).
	c) H factor (both for $\tau_E^{89p}$ and $\tau_E^{97L}$ scaling laws) as a function of edge pressure and input power, with colorbar indicating fusion gain.
	d) Fusion gain as a function of edge pressure and input power, with colorbar indicating the ratio of edge pressure over the \EPED-predicted H-mode pressure at each density.}
	\label{Fig:Q}
\end{figure}

Figure~\ref{Fig:Q} summarizes the fusion gain predictions for the 12 scenarios considered, as a function of the three controllable parameters.
When all three are plotted independently (Figure~\ref{Fig:Q}a), the fusion gain is seen to increase with edge density and edge temperature, as expected, and to either decrease or increase with input power, depending on the edge temperature (as discussed in Figure~\ref{Fig:stiff}).
However, when the edge pressure is used (Figure~\ref{Fig:Q}b), the fusion gain is seen to increase linearly with the pressure, organizing all results in roughly the same linear trend.
The effect of input power is seen most clearly at the highest edge pressure, a plasma in which operation with $8.3MW$ of heating power would result in $Q\approx1.2$ (\textit{breaking even}), while operation with $18.3MW$ would result in $Q\approx0.8$.
The reader is reminded that this work, focused on the core physics, does not make any claim about whether the edge pressure of $p_{\mathrm{edge}}\approx150kPa$ is achievable with only $8.3MW$ of input power. Further work is needed to relate these core predictions to the edge conditions.

Figures~\ref{Fig:Q}b), c) and d) have colorbars that indicate two metrics of importance to understand the validity of these predictions and the integration of core and edge physics.
H-factors (with respect to the $\tau_E^{89p}$ \cite{Yushmanov1999} and the $\tau_E^{97L}$ \cite{kaye_iter_1997} L-mode confinement scaling laws) in the vicinity of $H\approx1.2-1.3$ might be needed to achieve breakeven, which correspond to edge pressures $\sim35\%$ that of H-mode as predicted by \EPED \cite{Snyder2009}.
Generally, for the same power and boundary conditions, it is observed that $\tau_E^{97L}$ predicts higher ($\sim25\%$) H-factors than $\tau_E^{89p}$, which means that the later scaling law is more conservative and requires lower edge pressures for nominal confinement ($H=1.0$).

We must note that, even if L-mode H factors in excess of unity seem to be required for breakeven in SPARC according to the work presented here, the regression studies that lead to the construction of scaling laws often have large errors (usually $\sigma=\pm15\%$).
Furthermore, machines whose data was used to construct the scalings often achieve energy confinement levels higher (and lower) than the predicted scalings, owing to limitations of reproducing complex plasma dynamics with just a few global parameters.
Therefore, the results presented here should not be taken as overly pessimistic, as $H\approx1.2-1.3$ might be achievable in SPARC, particularly as we learn more about the edge physics and the control of core confinement in this new class of high-performance, burning-plasma machines.

\begin{figure}
	\centering
	\includegraphics[width=1.0\columnwidth]{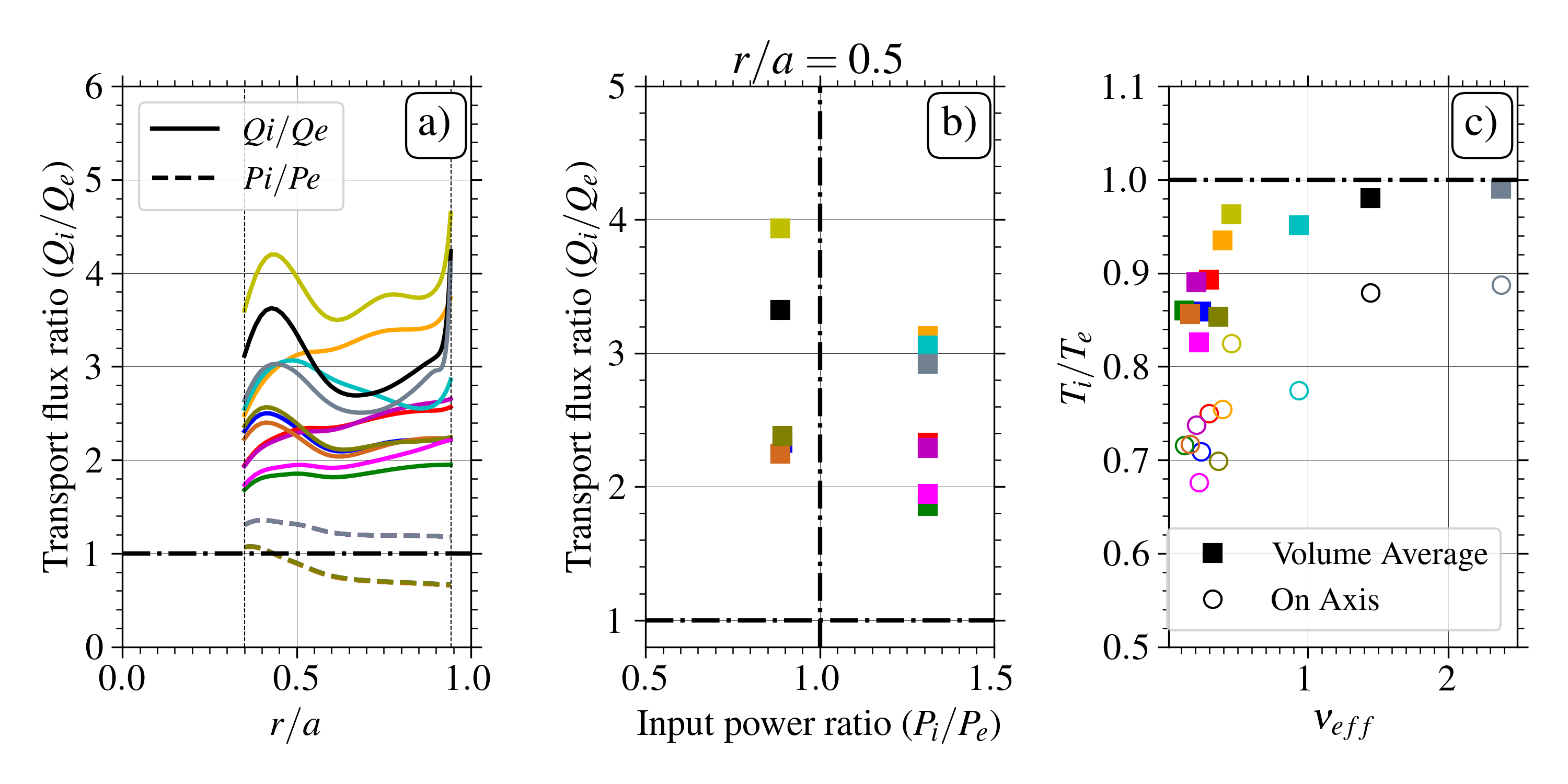}
	\caption{
        a) Radial profile of the ratio between ion and electron fluxes (transport flux in solid and input power flux in dashed).
        b) Transport flux ratio vs input power flux ratio at mid-radius ($r/a=0.5$).
        c) Ion to electron temperature ratio on-axis and volume averages, as a function of the effective collisionality.
    }
	\label{Fig:ratios}
\end{figure}

An important factor that affects the predicted performance and that can explain, to some level, the differences between the predictions presented here and those presented in, e.g., Refs.~\cite{Creely2020a,Rodriguez-Fernandez2022}, is the temperature ratio.
Figure~\ref{Fig:ratios}c shows the temperature ratio as a function of effective collisionality, $\nu_\mathrm{eff}=0.02\cdot \langle n_e\rangle_{20}\cdot R_{0,m}\cdot \langle T_e\rangle_{keV}^{-2}$, indicating that the ratio between volume-averaged temperatures is found to be $T_i/T_e=0.8-1.0$ and a strong function of collisionality.
Only those cases at high collisionality (those assumed to have low edge temperature and/or high edge density) display a temperature ratio close to unity, while the rest (and most) of cases have $T_i/T_e\approx0.85$.
This has an important effect when doing empirical modeling, as an operational point in POPCONs with the same confinement time and volume-averaged electron temperature will produce less fusion power if the actual ratio is below unity. Empirical scalings and POPCON modeling of burning plasmas often assume $T_i/T_e=1.0$, an assumption that needs to be revisited.

Interestingly, the steady-state ion to electron heat flux ratio remains high, above $Q_i/Q_e\gtrsim1.6$ at all locations for all cases studied here, as shown in Figure~\ref{Fig:ratios}a, regardless of whether there is more electron heating ($P_\mathrm{OH}+P_{\mathrm{ICRH},e}$) than ion heating ($P_{\mathrm{ICRH},i}$).
In fact, the heat flux ratio seems to be independent of the input power ratio, as shown in Figure~\ref{Fig:ratios}b, with electron-heating dominated plasmas having as high or even higher ion to electron heat flux ratio.
This is consistent with pilot plant case studies \cite{holland_development_2023} that pointed to the dominance of energy transport through the ions even in electron-heating dominated plasmas as a consequence of radiation, energy exchange and the ITG nature of the turbulence.
\referee{Even though the capability to include the turbulent energy exchange power as part of the energy balance has recently been implemented in \PORTALS \cite{rodriguez-fernandez_enhancing_2023}, the simulations presented here were performed prior to this addition and therefore do not account for the turbulent energy exchange power.
Motivated by recent results in low-collisionality, ITG-dominated plasmas \cite{kato_energy_2024}, the turbulent exchange power was evaluated with \CGYRO for a characteristic radial location ($r/a=0.75$) at low and high collisionalities for the SPARC L-mode database presented here.
The results indicate that, while at high collisionality the effect is small ($<6\%$ of the total exchange power), at low collisionality the turbulent exchange power can be as high as $23\%$ of the total exchange power.
This result motivates further studies to evaluate the effect of the turbulent exchange power in the energy balance, but it is left for future work.
}

\begin{figure}
	\centering
	\includegraphics[width=0.6\columnwidth]{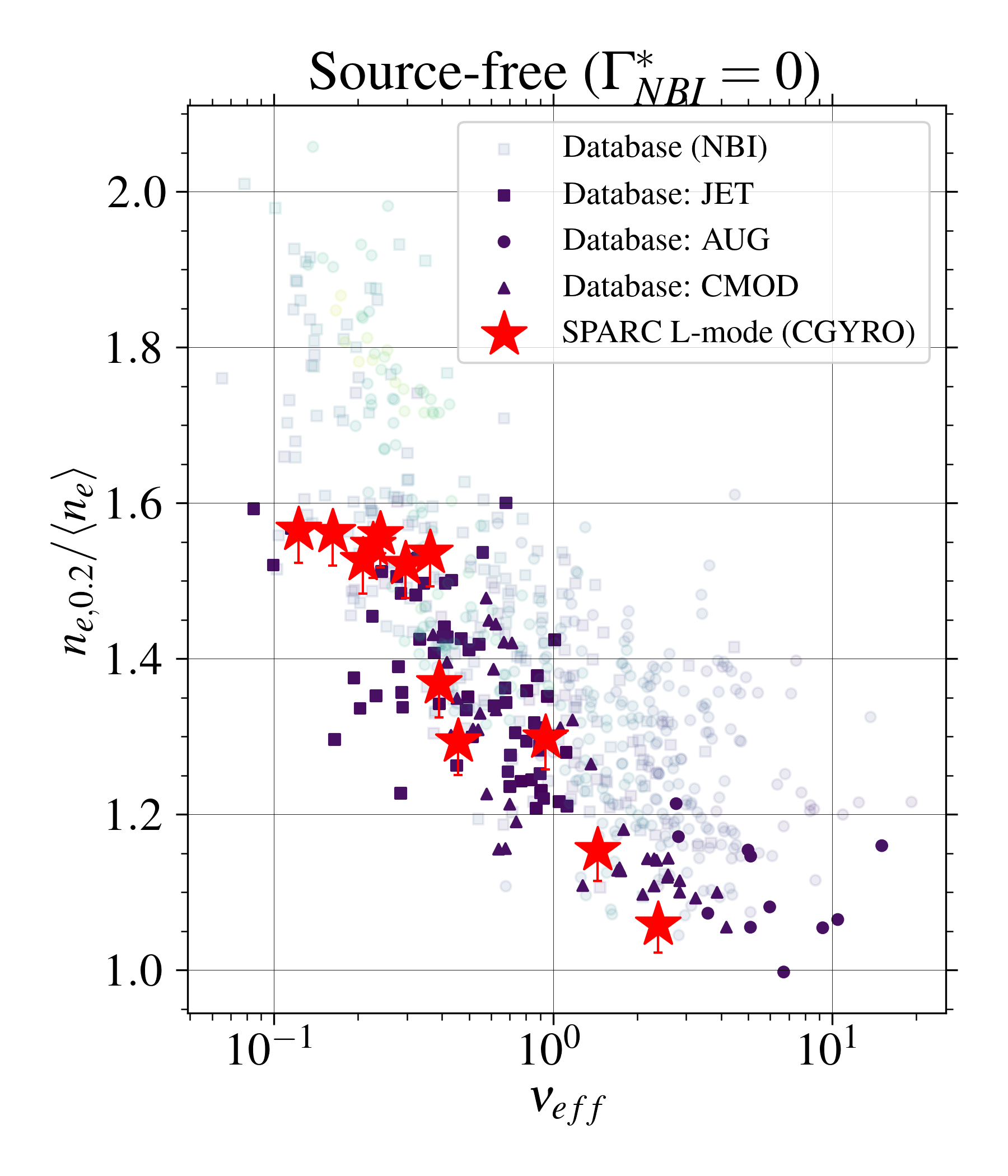}
	\caption{
        Electron density peaking as a function of the effective collisionality for the 12 cases studied here (red stars) and for a database of source-free Alcator C-Mod, ASDEX Upgrade and JET plasmas \cite{Angioni2007,Greenwald2007a} (purple).
        Cases from the database with core fueling from neutral beams have been shaded.
        Error bar on predicted peaking accounts for the edge density profile. The lower bound is obtained with a density profile flat in the $r/a=0.943-1.0$ region, and the upper bound is obtained with a density profile equal to zero (thus with a discontinuity at $r/a=0.943$) in that region.
    }
	\label{Fig:peaking}
\end{figure}

Density peaking is an important parameter to determine fusion performance, and in SPARC it will be determined by transport (balance of turbulent and neoclassical diffusion and convection), given the lack of significant core fueling.
Figure~\ref{Fig:peaking} shows the density peaking predicted in this work with \PORTALS-\CGYRO (which also accounts for neoclassical transport using the \NEO code) as a function of the effective collisionality. In this figure, the predictions are compared to the database of source-free plasmas in Alcator C-Mod, ASDEX Upgrade and JET that was used to derive the \textit{Angioni scaling} of density peaking \cite{Angioni2007,Greenwald2007a}.
Even though the database belonged to H-modes in the three machines, the density peaking predicted by \CGYRO and \NEO for this set of SPARC plasmas is found to be in remarkable agreement over a wide range of collisionalities (note that the x-axis is in logarithmic scale).

\subsection{On the effect of impurity content}
\label{sec:impurity}

Previous results correspond to the PRD nominal choice of $f_{DT}=0.85$ and $Z_\mathrm{eff}=1.5$, which is achieved in the simulations by assuming a ``lumped'' low-Z impurity that satisfies the quasineutrality and $Z_\mathrm{eff}$ constraints (accounting for the 5\%  ICRH minority concentration).
Here we explore the effect of an increased effective ion charge on the predictions, by either controllable (divertor impurity seeding) or uncontrollable sources.
For this study, we select a case at high input power ($P_\mathrm{in}=18.3MW$) and in middle of the $T-n$ database (case circled in Figure~\ref{Fig:database}b, with $T_\mathrm{edge}=1.6keV$ and $n_\mathrm{edge}=1.2\cdot10^{20}m^{-3}$).
We study two additional impurity levels: $Z_\mathrm{eff}=2.0$ ($f_{DT}=0.79$) and $Z_\mathrm{eff}=2.5$ ($f_{DT}=0.73$), which complement the nominal assumption of $Z_\mathrm{eff}=1.5$ ($f_{DT}=0.85$).
These higher impurity content plasmas are achieved by increasing the concentration of the ``lumped'' impurity until the desired $Z_\mathrm{eff}$ is reached.
Impurity charge ($Z=9$) and tungsten concentration ($f_W=1.5\cdot10^{-5}$) are assumed fixed.

\begin{figure}
	\centering
	\includegraphics[width=1.0\columnwidth]{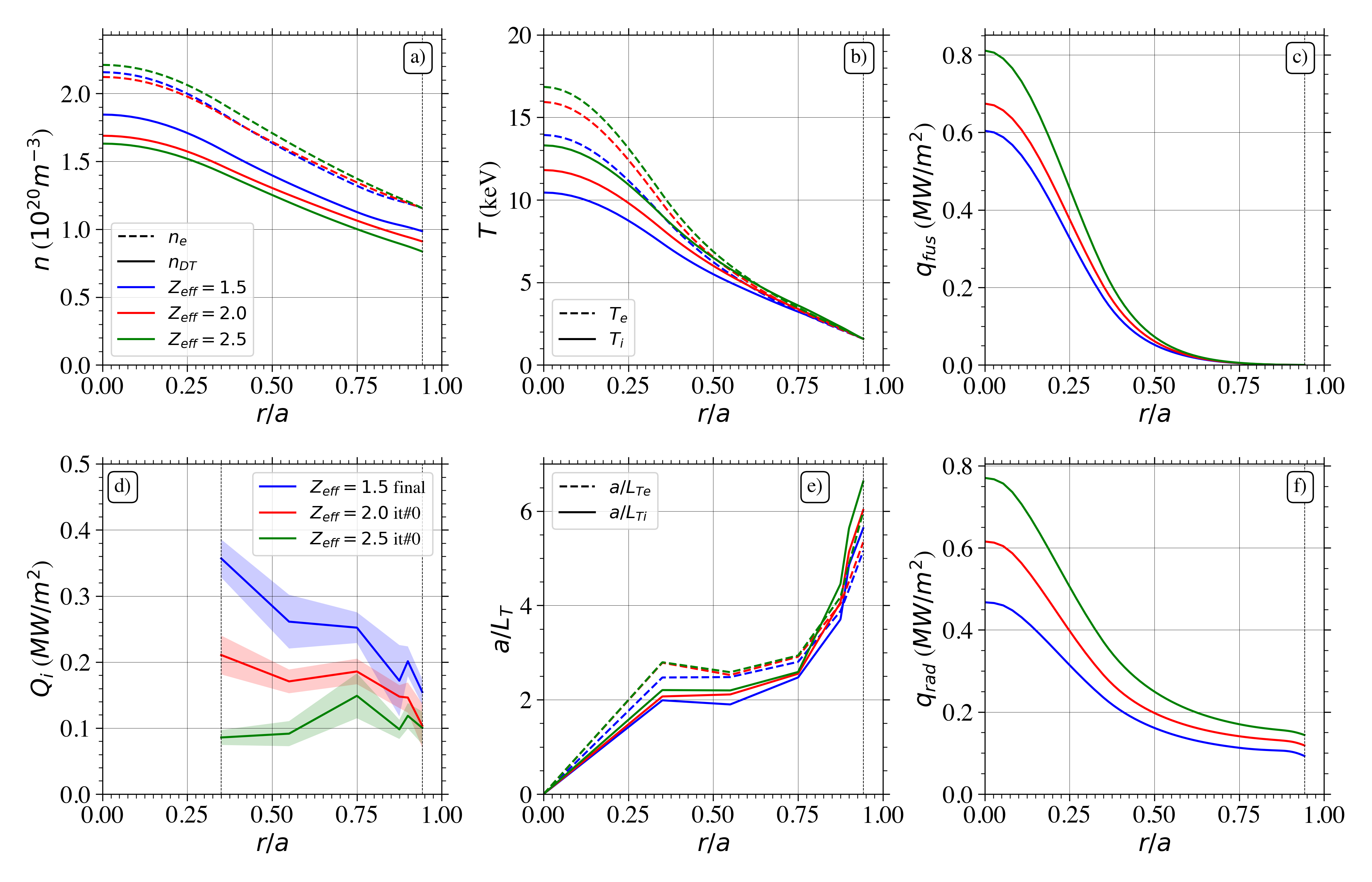}
	\caption{Study of the effect of impurity content on performance. a) Electron and fuel density profiles, b) electron and ion temperature profiles, c) fusion power density profile, d) ion energy flux comparison between converged nominal case and transient (iteration \#0) higher $Z_\mathrm{eff}$ cases, e) normalized logarithmic ion temperature gradient, and f) radiated power profile.
    }
	\label{Fig:zeff}
\end{figure}

Nonlinear gyrokinetic profile predictions, at constant edge conditions, display an increase in performance as the impurity content increases, as shown in Figure~\ref{Fig:zeff}.
Not only higher ion temperatures (Figure~\ref{Fig:zeff}b) are achieved via turbulence stabilization ---i.e. higher steady-state normalized logarithmic ion temperature gradients (Figure~\ref{Fig:zeff}e), for the same input power--- despite higher radiation losses (Figure~\ref{Fig:zeff}f), but also higher fusion power (Figure~\ref{Fig:zeff}c) despite the lower fuel density (Figure~\ref{Fig:zeff}a).
The nominal $Z_\mathrm{eff}=1.5$ case had $Q\sim0.29$, and increasing the impurity content led to $Q\sim0.33$ and $Q\sim0.4$ for $Z_\mathrm{eff}=2.0$ and $Z_\mathrm{eff}=2.5$ respectively, i.e. a $38\%$ increase in fusion gain from the nominal to the $Z_\mathrm{eff}=2.5$ case.
\referee{This increase in performance via impurity-induced suppression of core turbulence is consistent with past experimental and modeling studies \cite{mckee_impurity-induced_2000,McKee2000,ennever_effects_2015,hatch_gyrokinetic_2017,li_impurity_2019}.}

To understand the turbulence stabilization effect, Figure~\ref{Fig:zeff}d) shows a comparison between the steady-state ion energy flux for the nominal $Z_\mathrm{eff}=1.5$ case and the fluxes (coming from turbulent \CGYRO and neoclassical \NEO transport) that result in increasing the impurity content to $Z_\mathrm{eff}=2.0$ and $Z_\mathrm{eff}=2.5$.
As the target flux is very similar, the gradients need to peak more to achieve similar levels as in the $Z_\mathrm{eff}=1.5$ case (blue), resulting in the increased performance in flux-matched conditions.
For clarification, fluxes of the  $Z_\mathrm{eff}=2.0$ and $Z_\mathrm{eff}=2.5$ cases in Figure~\ref{Fig:zeff}d) are not in steady-state, but represent the iteration \#0 of the \PORTALS-\CGYRO process, which corresponds to the gradients of the nominal (converged) $Z_\mathrm{eff}=1.5$ case but with the differing impurity content.

These results are very promising and suggest that a path for higher performance may exist that also satisfies at the same time core-edge integration constraints (divertor heat flux limits and associated impurity seeding needs).
The further exploration of such operational scenarios and the investigation of whether the lower turbulent transport levels are due solely to the effect of the effective ion charge on the ion temperature critical gradient or on dilution effects (which can lead to interesting optimization studies of the puffed impurity of choice) will be subject of future work.

\section{Quasilinear gyro-fluid results}
\label{sec:PORTALStglf}

The \TGLF quasilinear model for core turbulence has been tested against the database of 528 nonlinear gyrokinetic results that were presented in this work.
With a much reduced computational cost, \TGLF estimates transport fluxes by solving a linear system of gyro-fluid equations \cite{Staebler2005} as an eigenvalue problem and uses a saturation rule \cite{Staebler2007} to calculate fluxes from the growth rates and real frequencies of unstable micro-instabilities, together with the eigenmode solution and associated quasilinear weights.
The saturation rule in \TGLF has evolved over the years, as the community has gathered more understanding of the dynamics of turbulence in the plasma core.
Here we test the three latest saturation rules, SAT1 \cite{Staebler2017,Staebler2020}, SAT2 \cite{Staebler2021}  and SAT3 \cite{dudding_new_2022}, with electromagnetic effects 
($\delta \phi$, $\delta A_{\parallel}$, and $\delta B_{\parallel}$) and only capturing long-wavelength turbulence ($k_\theta\rho_s\leq1.22$) for consistency with the \CGYRO simulations.
As a reminder, the SAT1, SAT2 and SAT3 saturation rules were originally constructed to reproduce the cross-scale turbulence coupling, the 3D dependence (poloidal angle, radial and poloidal wavenumber) of the saturated intensity and the isotope effect, respectively, with their regression coefficients fitted with databases of a few tens of nonlinear \CGYRO simulations.

\begin{figure}
	\centering
	\includegraphics[width=1.0\columnwidth]{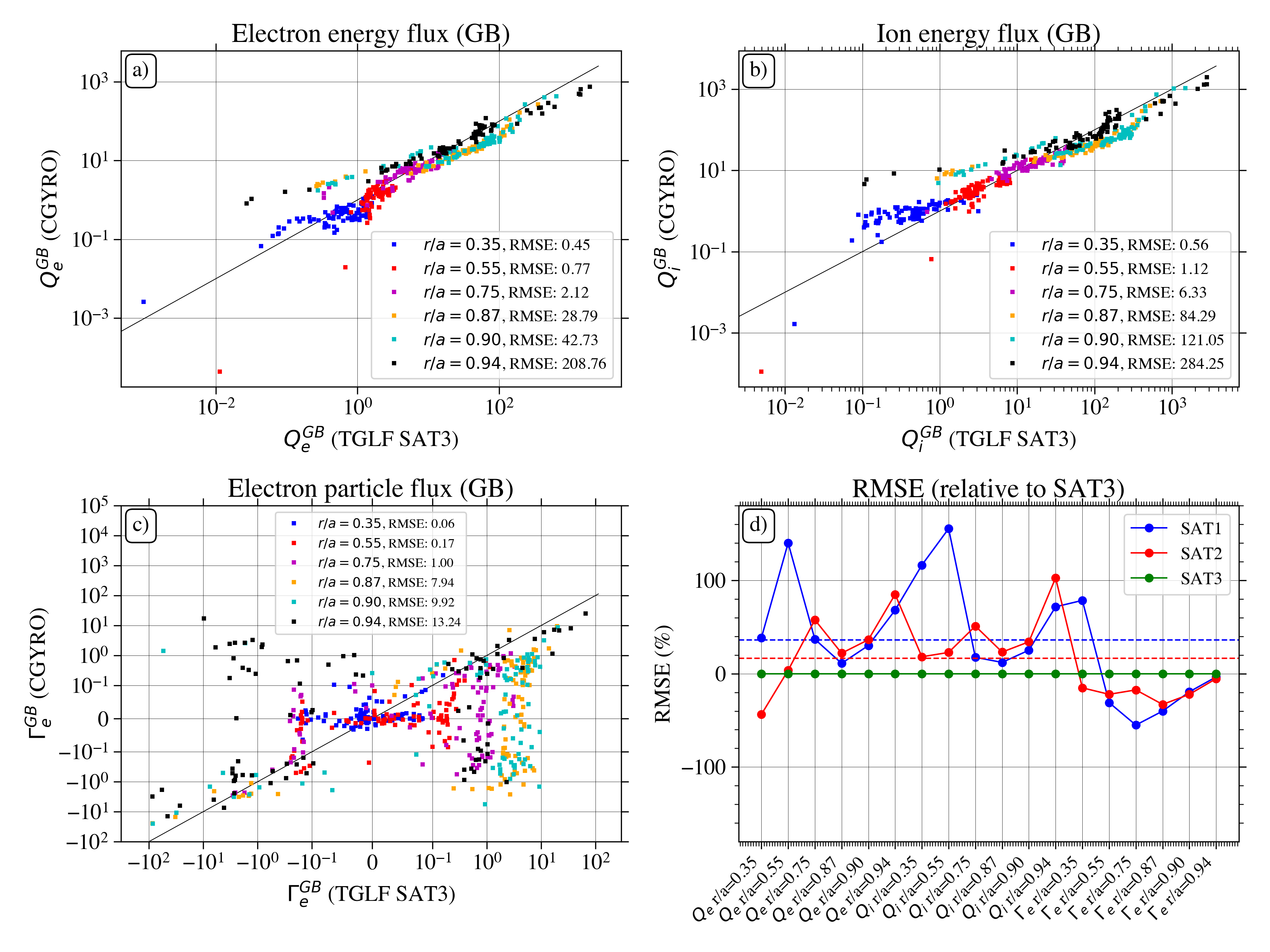}
	\caption{
        Validation of \TGLF saturation rules against nonlinear \CGYRO for the database of 14 scenarios (12 nominal scenarios + 2 variations of $Z_\mathrm{eff}$), resulting in 528 local nonlinear \CGYRO simulations to be compared to \TGLF. a) Electron energy, b) ion energy and c) electron particle fluxes at the 6 flux surfaces for \TGLF SAT3 and nonlinear \CGYRO.
		d) Relative change in root mean square error of the three transport fluxes at each location when changing the saturation rule from SAT3 to SAT1 and SAT2.
        Horizontal dashed lines indicate the average (radius and channel) relative change in root mean square error.
		For those points above zero, SAT3 performs better.
    }
	\label{Fig:tglf_std}
\end{figure}

Figure~\ref{Fig:tglf_std} shows a comparison of \CGYRO and \TGLF predictions for each of the three transport fluxes under consideration (electron energy, ion energy and electron particle fluxes) at the 6 flux surfaces considered in this work, totalling 528 local simulations.
Remarkably, \TGLF is capable of reproducing the transport fluxes well, spanning over four orders of magnitude. This is in spite of the fact that \TGLF is fitted to only a few tens of \CGYRO simulations per saturation rule, in different parameter spaces than this database of SPARC plasmas, and with just $\sim$10 CPU-seconds per evaluation (vs $\sim$200 GPU-hours for nonlinear \CGYRO).
Generally, it is observed that SAT3 performs the best among the saturation rules tested in this work. As seen in Figure~\ref{Fig:tglf_std}d, SAT3 performs \referee{on} average better than SAT1 and SAT2.
The strongest differences between \CGYRO and \TGLF SAT3 are observed at the edge, where \TGLF is generally seen to overpredict electron and ion energy fluxes.
Unfortunately, SAT3 performs the worst of the three at reproducing particle fluxes.
Interestingly, SAT2 is found to perform \referee{worse} (highest root mean square error) than SAT1 and SAT3 at reproducing energy fluxes at the plasma edge (positions $r/a=0.9$ and $0.943$).
This is somewhat in contrast to the results presented in Refs.~\cite{Angioni2022,angioni_dependence_2023}, where SAT2 was found to be in good agreement with ASDEX Upgrade experiments and capable of reproducing the parametric dependencies of the L-mode scaling law well.

\begin{figure}
	\centering
	\includegraphics[width=1.0\columnwidth]{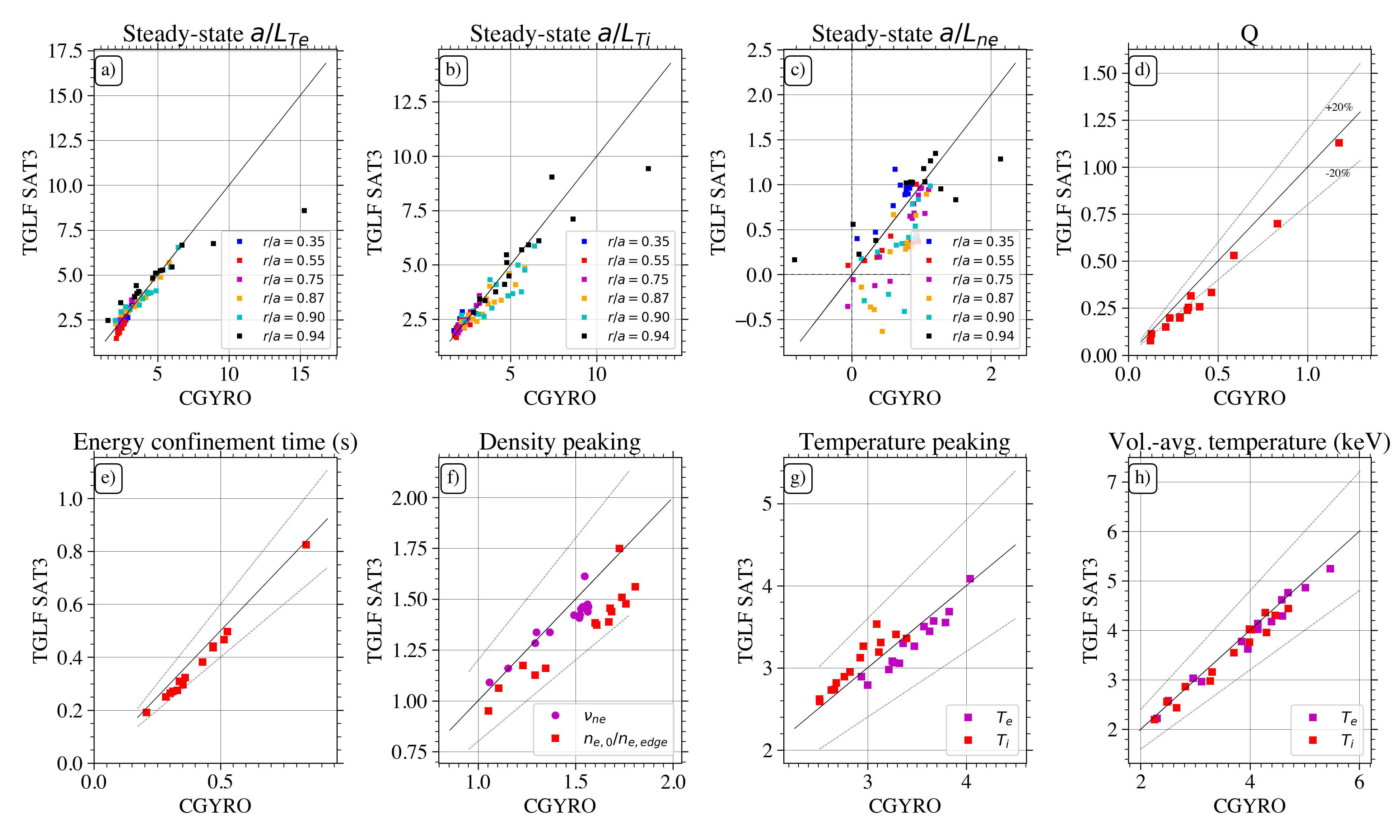}
	\caption{
		Validation of \TGLF SAT3 against nonlinear \CGYRO in multi-channel, flux-matched conditions for both models.
		Plots provide comparisons between \CGYRO (x-axis) and \TGLF SAT3 (y-axis) of the following quantities:
		a), b) and c) local normalized logarithmic gradients of the electron temperature, ion temperature and electron density in steady state conditions, d) fusion gain, e) energy confinement time, \referee{f)} density peaking (both standard definition, $n_{e,0.2}/\langle n_e\rangle$, and $n_{e,0}/n_{e,edge}$), g) electron and ion temperature peaking, and h) volume-averaged electron and ion temperatures.
    }
	\label{Fig:tglf_pp}
\end{figure}

A comparison of gradient-driven transport fluxes such as the one just described in Figure~\ref{Fig:tglf_std} is not entirely fair, as the critical gradient and stiff nature of turbulence in the core of tokamak plasmas may result in very different predicted fluxes for very similar gradients.
Therefore, a flux-matched comparison of predicted kinetic profiles can provide more important insights.
\PORTALS-\TGLF (SAT3) was run for all cases in the database, and the results of comparing the predicted profiles and performance with \PORTALS-\CGYRO are shown in Figure~\ref{Fig:tglf_pp}.

\begin{figure}
	\centering
	\includegraphics[width=1.0\columnwidth]{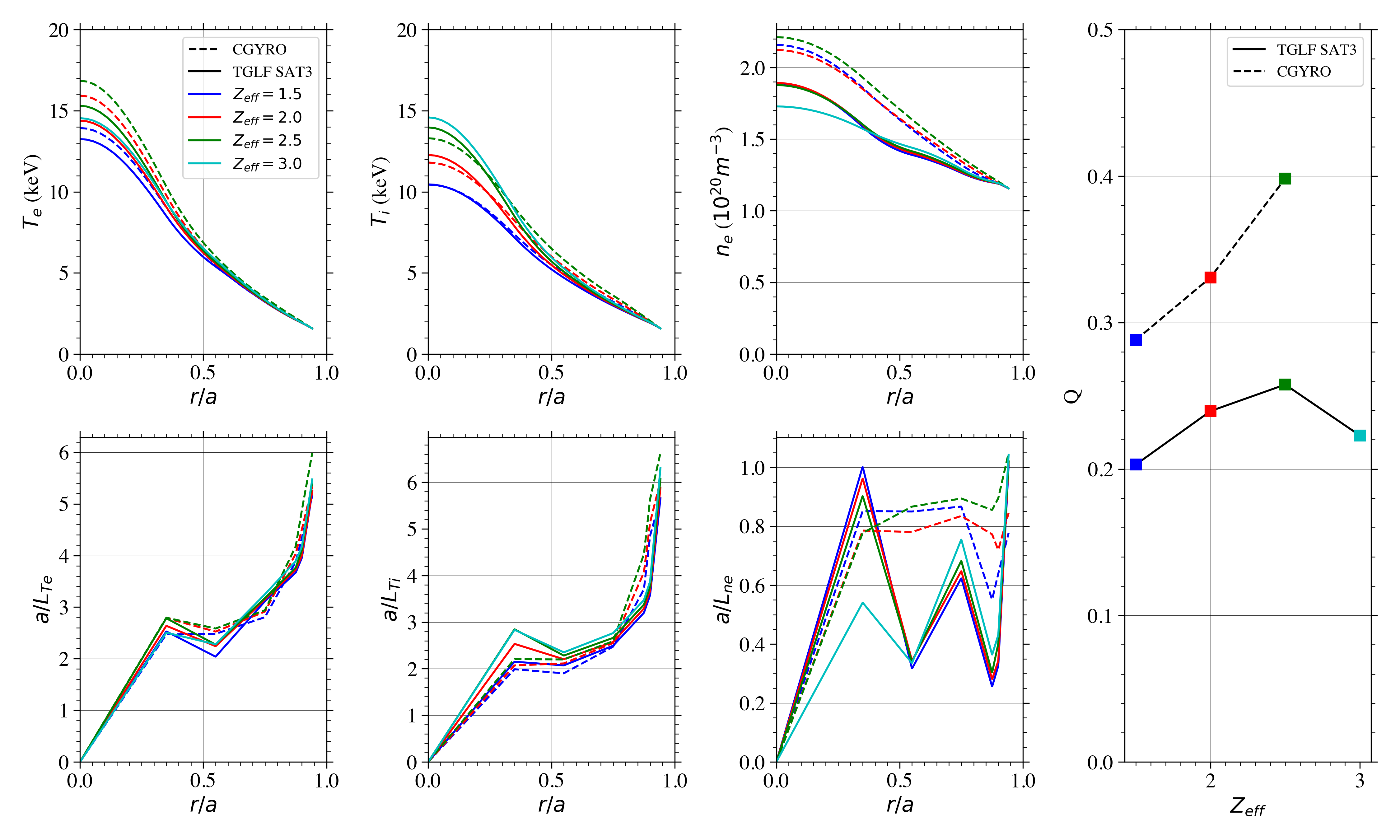}
	\caption{
        Study of the validity of \TGLF to predict the effect of impurity content on performance: profiles of electron temperature, ion temperature and electron density profiles and their respective logarithmic gradients. Fusion gain vs $Z_\mathrm{eff}$ is also shown. $Z_\mathrm{eff}=3.0$ case is only available with \TGLF as of the time of writing.
    }
	\label{Fig:tglf_pp_zeff}
\end{figure}

Generally, \TGLF SAT3 is seen to reproduce the predicted temperature profiles well (Figures~\ref{Fig:tglf_pp}a and b), with the exception of the very edge, $r/a=0.943$, where \CGYRO predicted higher gradients for both electron and ion temperatures. This is consistent with the results in Figure~\ref{Fig:tglf_std} that showed \TGLF SAT3 overpredicting the transport fluxes at the edge, which leads to lower gradients in steady-state conditions when the plasma has a non-stiff edge (low density and temperature boundary conditions).
In global terms, ion temperature peaking is seen to be underpredicted by \TGLF SAT3, while electron temperature peaking is overpredicted (Figure~\ref{Fig:tglf_pp}g). The volume-averaged temperatures are seen to be slightly underpredicted by \TGLF SAT3 (Figure~\ref{Fig:tglf_pp}h), which is consistent with the overprediction of the transport fluxes at the edge.
As per the electron density gradients (Figure~\ref{Fig:tglf_pp}c), the agreement is not as good as for the temperature gradients, with more scattered behavior and the appearance of hollowed profiles with \TGLF SAT3, particularly at $r/a=0.9$ and $0.87$.
This leads generally to an underprediction of the density peaking with \TGLF(Figure~\ref{Fig:tglf_pp}f), consistent with other studies with quasilinear profile predictions \cite{rodriguez-fernandez_nonlinear_2022}.
Example density profiles showing disagreement (lower peaking with the quasilinear model) between \CGYRO and \TGLF are depicted in Figure~\ref{Fig:tglf_pp_zeff}.

In terms of performance, energy confinement time is underpredicted by \TGLF, a consequence of the underprediction of both the volume averaged temperatures and density peaking.
The lower temperatures and densities lead to lower fusion power, for which \CGYRO predicts generally higher performance.
These results are, however, encouraging, as \TGLF SAT3 is capable of predicting performance reasonably well at a tiny fraction of the computational expense, which will allow the further exploration of the parameter space and optimization studies at much reduced cost.

\TGLF SAT3 is capable of predicting the increase of performance with the increase in impurity content, as shown in Figure~\ref{Fig:tglf_pp_zeff}.
In \CGYRO, it was observed that $Q$ increased by $\sim38\%$ from the nominal $Z_\mathrm{eff}=1.5$ case to the $Z_\mathrm{eff}=2.5$ case, while in \TGLF $Q$ increased by $\sim30\%$.
However, the physics reasons behind the improvement in performance are different.
In \TGLF, the change in edge temperature gradients is not as strong as in \CGYRO, and the improvement in performance is mostly due to the increase in near-axis ($r/a=0.35$) temperature gradients.
An additional case with $Z_\mathrm{eff}=3.0$ was also studied with \TGLF, and it is seen to predict a roll-over in performance, which is consistent with the dilution of fusion fuel becoming too strong to be compensated by the turbulence stabilization, as $f_{DT}=0.67$ for $Z_\mathrm{eff}=3.0$.

It is important to note that predictions with \TGLF spanning to short wavelengths (\TGLF default $k_\theta\rho_s\leq24.0$ instead of $k_\theta\rho_s\leq1.22$ used here for consistency with \CGYRO) did show differences in standalone fluxes and in kinetic profile predictions, particularly with electron energy and particle fluxes having non-negligible contributions at $k_\theta\rho_s\sim2-3$ at the larger radii, even in situations with $\max\left(\gamma/k_\theta\rho_s\right)_{high-k} < \max\left(\gamma/k_\theta\rho_s\right)_{low-k}$, a zonal flow mixing criterion usually employed to assess the relative influence of multi-scale effects \cite{Creely2019}.
While this could suggest the need to run nonlinear gyrokinetic simulations that also include shorter wavelengths than used in this study \referee{(to capture mid and high wavenumber TEM and ETG modes, and cross-scale interactions)}, further work is needed to verify the capabilities of \TGLF to capture cross-scale interactions and mid-wavenumber flux contributions correctly, for which very limited studies exist.
Due to the very large computational resources required for such an exercise, this is out of the scope of this work but will be the subject of future investigations.

\section{Discussion}
\label{sec:Discussion}

Sections~\ref{sec:PORTALScgyro} and \ref{sec:PORTALStglf} have presented and discussed the results of the ion-scale nonlinear gyrokinetic and quasilinear gyrofluid modeling of SPARC core plasmas when four parameters are varied: the edge ($r/a=0.943$) density, the edge temperature, the input power and the impurity content.
While there are still unknowns and the assumed edge conditions are not fully understood, the results presented here provide a few take-away lessons that will be useful for the design of experiments and operation of SPARC and other future burning-plasma machines.

It is found that L-modes characterized by energy confinement times in line with empirical scaling laws (i.e. $H\sim1.0$; where $H$ is the confinement factor relative to empirical L-mode confinement time) in high-performance plasmas require moderate ($\sim$35\% that of H-mode) edge pressures.
In these hot, dense conditions, turbulence is very stiff, as a consequence of the high gyro-Bohm unit and dominance of ITG turbulence, generally expected in reactor-like plasmas \cite{holland_development_2023}.
As such, these plasmas behave more closely to present-day H-modes than to L-modes, which could also explain the very good agreement with the database of H-mode density peakings.
In conditions closer to $H\sim1.0$, the normalized logarithmic temperature gradient profiles become radially flat and stiffer, requiring a high edge pressure to maintain the energy confinement time as predicted by the scalings.
In general, this opens the question of the validity of L-mode confinement scaling laws in high-temperature, high-density plasmas (i.e. reactor-like and burning plasmas), where the core behaves more similarly to H-modes.
This apparent contradiction may be clarified by looking at core gyrokinetic turbulent transport as a dimensionless, local process, where normalized quantities such as logarithmic gradients normalized to the plasma minor radius, collisionality or temperature ratios are the ones that determine the gyro-Bohm normalized turbulent transport levels.
In this sense, as long as the assumptions of flux-tube gyrokinetic theory (well attained in burning plasmas) are valid, the core transport dynamics are independent of the edge conditions.
If L-modes in high-performing devices like SPARC are empirically found to achieve high core temperatures and densities, the behavior of the core may be more similar to H-modes, where the high temperature is achieved by means of an edge transport barrier.
The lack of neutral beam injectors and limited flexibility to control the current density profile in SPARC may mean that the formation of internal transport barriers should not be, generally, expected.
\referee{It is also important to note that the increased stiffness found in these high-performance ``L-mode'' plasmas, which potentially separates them from present-day machines, also implies that the edge transport similarity assumed when using the Alcator C-Mod database of edge gradient profiles needs to be revisited.
Future work will include the characterization of the edge transport in SPARC L-modes via multi-machine databases that span a wide range of edge collisionalities and input powers.	
}

The requirement of high edge pressure may be a challenge for the integration of core performance and edge and divertor constraints, particularly if a transport-regulated edge barrier, $r/a=0.943-1.0$, is needed.
As a result, the sustainment of a high enough core energy confinement time in ``L-mode'' conditions may still require high input power as a result of the high magnetic field that contributes to a low $E\times B$ shearing rate \cite{angioni_dependence_2023}. 
This high power to be transported through $r/a=0.943-1.0$ to sustain the high gradient would be challenging for managing a reasonable divertor heat flux, particularly during early operations, when the divertor conditions and flux mitigation will still be somewhat unknown and solutions to the heat flux and loading challenges will still be under development.
Given the present uncertainties and risks that we will only be able to retire after first experiments have been performed, the SPARC team is devising contingency plans that include potentially pivoting to H-mode operation (including at reduced magnetic field, $8T$ \cite{rodriguezfernandez_aps21}) if needed in order to get the $Q>1$ milestone.

The attainment of a high edge pressure may be achieved by means of access to I-mode regimes \cite{hughes_fec23}, which have potential advantages to H-modes, such as ELM avoidance, low impurity accumulation and better density control.
\referee{I-mode edge profiles were almost certainly the main contributors to the edge profiles for the high edge temperatures ($T_{edge}>0.6keV$) shown in the database of Alcator C-Mod plasmas in Figure~\ref{Fig:bc}. }
The rather large ion to electron heat flux ratio seen in Figure~\ref{Fig:ratios} (even at low Greenwald fractions, below what could be estimated to be the transition to the low-density branch for mode transitions \cite{Ryter2014}) may also be a sign of the potential for I-mode-like regimes, as there is evidence for a critical ion heat flux required for L to I transitions \cite{ryter_i-mode_2016}.
The potential for I-mode access is being examined and will be the subject of a follow-on publication.
The high $Q_i/Q_e$ ratio also suggests that $T_i>T_e$ may be observed at the separatrix, which could further lead to higher fusion power at the same edge pressure.
However, its impact at the boundary condition for the core simulations ($r/a=0.943$) has not been explored yet and it will be part of future work.
\referee{The modeling of transport from separatrix to the core, coupling fueling, edge $E\times B$ shear and integration with scrape-off layer and divertor physics models (a truly self-consistent integrated model) will also be part of future work.}

Interestingly, this work has found that increasing the impurity content in these plasmas, even if the fusion fuel becomes more diluted, may lead to higher fusion power and performance, under the assumption of a fixed edge boundary condition.
This is consistent with previous gyrokinetic modeling and experimental validation work \cite{ennever_effects_2015}.
This work so far has only explored the effect of increasing the effective ion charge with a fixed low-Z, ``lumped'' impurity, but future work will explore the effect of other impurity choices and their concentration.
This is very encouraging, as lower edge pressures may be allowable if higher fusion power is attained by means of core turbulence stabilization via higher impurity content.
Other potential phenomena, such as the effect of fast ions on turbulence \cite{siena_predictions_2023}, have not been addressed in this work, which could also favor high-performance cores with lower edge pressures.
Toroidal rotation has also been assumed to be zero throughout the core, as well as the particle fueling. Both effects are generally expected to increase performance (the effect of rotation on standalone turbulence fluxes was explored in Ref.~\cite{Howard2021} for the SPARC PRD scenario), but how much of an increase still remains an open question, until reliable first-principles or physics-based models of edge particle fueling and residual momentum stresses are available in a fully predict-first, physics-based  fashion.

\section{Conclusions}
\label{sec:Conclusions}

This work has presented a high-fidelity core ($r/a<0.943$) transport study of, arguably, unprecedented fidelity in magnetically confined plasmas. Thanks to surrogate-acceleration of transport solvers enabled by \PORTALS, nonlinear gyrokinetics can now be used to predict operational scenarios in future burning plasmas and in this work a large database of scenarios was predicted at a moderate computational expense.

Results of the parametric exploration of SPARC early-campaign scenarios project near-breakeven conditions in L-mode-like plasmas. Breakeven is predicted when the edge pressure becomes $\sim35\%$ that of H-mode, but the actual fusion gain will be determined by how much input power is required to sustain such edge pressures.
Higher impurity content is, somewhat unexpectedly, predicted to increase fusion power, opening an interesting path for optimization and transport physics investigations.
Enabled by an extensive database of 528 local turbulence simulations, quasilinear results with \TGLF show promise for the use of reduced models to widely explore the parameter space because, although some differences appear (particularly in the density profile predictions and edge temperature gradients), \TGLF is able to capture the parametric dependencies well, together with the overall plasma performance, at a fraction of the cost.

This work has pointed out the need to understand better the edge conditions in L-mode plasmas, as well as the validity of energy confinement scaling laws to predict the performance of high-temperature and high-density L-modes.
The first plasmas in upcoming machines such as SPARC and ITER will be L-mode-like. Furthermore, the challenge of core-edge integration, ELM mitigation and divertor heat flux control has recently led fusion researchers to explore the potential of non-H-mode operation directly in fusion power plant studies (e.g. \cite{Sorbom2015,frank_radiative_2022,wilson_aps23}).
It is only by the careful validation \cite{White2019} of predictive simulations against experimental data that we will be able to understand the physics of these plasmas and to optimize the performance of such devices \cite{fable_high-confinement_2021}.
\referee{More importantly, L-mode operation in high-performance devices could be more similar to H-mode operation than to present-day L-modes, consequence of the high gyro-Bohm transport in high temperature, high density core plasmas.
This could have important implications for the design and operation of future fusion power plants.
The validation of this modeling insight in the first campaigns of SPARC will certainly be a key milestone in the path to economic fusion energy.
}

The upcoming years, as we witness SPARC's first experimental operations, promise not only a test-bed for our predictions but also a fertile ground for learning and improving our predictive capabilities.
This cycle of prediction, validation, and refinement is crucial as we navigate from achieving burning plasma conditions to envisioning fusion power plants.
The insights from validating high-fidelity simulations such as the ones presented in this work against real-world data will be invaluable, guiding the optimization of future fusion devices.
As we refine these models, their improved predictive power will be instrumental in the design, operation, and optimization of fusion power plants, marking a significant leap towards a future powered by clean, sustainable fusion energy.

\section*{Acknowledgments}

The authors would like to thank M. Greenwald for the experimental data used in Figure~\ref{Fig:peaking}.
This work was funded by Commonwealth Fusion Systems under RPP020, and used resources of the National Energy Research Scientific Computing Center (NERSC), a US Department of Energy Office of Science User Facility located at Lawrence Berkeley National Laboratory, operated under Contract No. DE-AC02-05CH11231, for the \PORTALS-\CGYRO simulations (\CGYRO from \texttt{main} branch with hash \texttt{016ac47}, \PORTALS version 1.0.0).
Clusters hosted at the Massachusetts Green High Performance Computing Center were used to perform the \PORTALS-\TGLF simulations (\TGLF from \texttt{main} branch with hash \texttt{037b2f0}).
Alcator C-Mod data used in this work were obtained under DOE Award DE-FC02-99ER54512.
ChatGPT-4 and Microsoft Copilot were used to enhance parts of the manuscript for clarity and coherence.

\bibliography{References/references_fixed.bib,References/references_extra.bib}


\begin{thebibliography}{73}
\ifx \bisbn   \undefined \def \bisbn  #1{ISBN #1}\fi
\ifx \binits  \undefined \def \binits#1{#1}\fi
\ifx \bauthor  \undefined \def \bauthor#1{#1}\fi
\ifx \batitle  \undefined \def \batitle#1{#1}\fi
\ifx \bjtitle  \undefined \def \bjtitle#1{#1}\fi
\ifx \bvolume  \undefined \def \bvolume#1{\textbf{#1}}\fi
\ifx \byear  \undefined \def \byear#1{#1}\fi
\ifx \bissue  \undefined \def \bissue#1{#1}\fi
\ifx \bfpage  \undefined \def \bfpage#1{#1}\fi
\ifx \blpage  \undefined \def \blpage #1{#1}\fi
\ifx \burl  \undefined \def \burl#1{\textsf{#1}}\fi
\ifx \doiurl  \undefined \def \doiurl#1{\url{https://doi.org/#1}}\fi
\ifx \betal  \undefined \def \betal{\textit{et al.}}\fi
\ifx \binstitute  \undefined \def \binstitute#1{#1}\fi
\ifx \binstitutionaled  \undefined \def \binstitutionaled#1{#1}\fi
\ifx \bctitle  \undefined \def \bctitle#1{#1}\fi
\ifx \beditor  \undefined \def \beditor#1{#1}\fi
\ifx \bpublisher  \undefined \def \bpublisher#1{#1}\fi
\ifx \bbtitle  \undefined \def \bbtitle#1{#1}\fi
\ifx \bedition  \undefined \def \bedition#1{#1}\fi
\ifx \bseriesno  \undefined \def \bseriesno#1{#1}\fi
\ifx \blocation  \undefined \def \blocation#1{#1}\fi
\ifx \bsertitle  \undefined \def \bsertitle#1{#1}\fi
\ifx \bsnm \undefined \def \bsnm#1{#1}\fi
\ifx \bsuffix \undefined \def \bsuffix#1{#1}\fi
\ifx \bparticle \undefined \def \bparticle#1{#1}\fi
\ifx \barticle \undefined \def \barticle#1{#1}\fi
\bibcommenthead
\ifx \bconfdate \undefined \def \bconfdate #1{#1}\fi
\ifx \botherref \undefined \def \botherref #1{#1}\fi
\ifx \url \undefined \def \url#1{\textsf{#1}}\fi
\ifx \bchapter \undefined \def \bchapter#1{#1}\fi
\ifx \bbook \undefined \def \bbook#1{#1}\fi
\ifx \bcomment \undefined \def \bcomment#1{#1}\fi
\ifx \oauthor \undefined \def \oauthor#1{#1}\fi
\ifx \citeauthoryear \undefined \def \citeauthoryear#1{#1}\fi
\ifx \endbibitem  \undefined \def \endbibitem {}\fi
\ifx \bconflocation  \undefined \def \bconflocation#1{#1}\fi
\ifx \arxivurl  \undefined \def \arxivurl#1{\textsf{#1}}\fi
\csname PreBibitemsHook\endcsname

\bibitem{rodriguez-fernandez_enhancing_2023}
\begin{botherref}
\oauthor{\bsnm{Rodriguez-Fernandez}, \binits{P.}},
\oauthor{\bsnm{Howard}, \binits{N.T.}},
\oauthor{\bsnm{Saltzman}, \binits{A.}},
\oauthor{\bsnm{Kantamneni}, \binits{S.}},
\oauthor{\bsnm{Candy}, \binits{J.}},
\oauthor{\bsnm{Holland}, \binits{C.}},
\oauthor{\bsnm{Balandat}, \binits{M.}},
\oauthor{\bsnm{Ament}, \binits{S.}},
\oauthor{\bsnm{White}, \binits{A.E.}}:
Enhancing predictive capabilities in fusion burning plasmas through surrogate-based optimization in core transport solvers.
arXiv.
arXiv:2312.12610 [physics]
(2023).
\url{http://arxiv.org/abs/2312.12610}
2023-12-21
\end{botherref}
\endbibitem

\bibitem{Doyle2007}
\begin{barticle}
\bauthor{\bsnm{Doyle}, \binits{E.J.}},
\bauthor{\bsnm{Houlberg}, \binits{W.A.}},
\bauthor{\bsnm{Kamada}, \binits{Y.}},
\bauthor{\bsnm{Mukhovatov}, \binits{V.}},
\bauthor{\bsnm{Osborne}, \binits{T.H.}},
\bauthor{\bsnm{Polevoi}, \binits{A.}},
\bauthor{\bsnm{Bateman}, \binits{G.}},
\bauthor{\bsnm{Connor}, \binits{J.W.}},
\bauthor{\bsnm{Cordey}, \binits{J.G.}},
\bauthor{\bsnm{Fujita}, \binits{T.}},
\bauthor{\bsnm{Garbet}, \binits{X.}},
\bauthor{\bsnm{Hahm}, \binits{T.S.}},
\bauthor{\bsnm{Horton}, \binits{L.D.}},
\bauthor{\bsnm{Hubbard}, \binits{A.E.}},
\bauthor{\bsnm{Imbeaux}, \binits{F.}}, \betal:
\batitle{Progress in the {ITER} {Physics} {Basis} {Chapter} 2: {Plasma} confinement and transport}.
\bjtitle{Nuclear Fusion}
\bvolume{47},
\bfpage{18}
(\byear{2007}).
\doiurl{10.1088/0029-5515/47/6/S02}
\end{barticle}
\endbibitem

\bibitem{Creely2020a}
\begin{botherref}
\oauthor{\bsnm{Creely}, \binits{A.J.}},
\oauthor{\bsnm{Greenwald}, \binits{M.J.}},
\oauthor{\bsnm{Ballinger}, \binits{S.B.}},
\oauthor{\bsnm{Brunner}, \binits{D.}},
\oauthor{\bsnm{Canik}, \binits{J.}},
\oauthor{\bsnm{Doody}, \binits{J.}},
\oauthor{\bsnm{Flp}, \binits{T.}},
\oauthor{\bsnm{Garnier}, \binits{D.T.}},
\oauthor{\bsnm{Granetz}, \binits{R.}},
\oauthor{\bsnm{Gray}, \binits{T.K.}},
\oauthor{\bsnm{Holland}, \binits{C.}},
\oauthor{\bsnm{Howard}, \binits{N.T.}},
\oauthor{\bsnm{Hughes}, \binits{J.W.}},
\oauthor{\bsnm{Irby}, \binits{J.H.}},
\oauthor{\bsnm{Izzo}, \binits{V.A.}}, et al.:
Overview of the {SPARC} tokamak.
Journal of Plasma Physics
\textbf{86}(5).
\doiurl{10.1017/S0022377820001257}.
2021-07-12
\end{botherref}
\endbibitem

\bibitem{Rodriguez-Fernandez2022}
\begin{barticle}
\bauthor{\bsnm{Rodriguez-Fernandez}, \binits{P.}},
\bauthor{\bsnm{Creely}, \binits{A.J.}},
\bauthor{\bsnm{Greenwald}, \binits{M.J.}},
\bauthor{\bsnm{Brunner}, \binits{D.}},
\bauthor{\bsnm{Ballinger}, \binits{S.B.}},
\bauthor{\bsnm{Chrobak}, \binits{C.P.}},
\bauthor{\bsnm{Garnier}, \binits{D.T.}},
\bauthor{\bsnm{Granetz}, \binits{R.}},
\bauthor{\bsnm{Hartwig}, \binits{Z.S.}},
\bauthor{\bsnm{Howard}, \binits{N.T.}},
\bauthor{\bsnm{Hughes}, \binits{J.W.}},
\bauthor{\bsnm{Irby}, \binits{J.H.}},
\bauthor{\bsnm{Izzo}, \binits{V.A.}},
\bauthor{\bsnm{Kuang}, \binits{A.Q.}},
\bauthor{\bsnm{Lin}, \binits{Y.}}, \betal:
\batitle{Overview of the {SPARC} physics basis towards the exploration of burning-plasma regimes in high-field, compact tokamaks}.
\bjtitle{Nuclear Fusion}
\bvolume{62}(\bissue{4}),
\bfpage{042003}
(\byear{2022}).
\doiurl{10.1088/1741-4326/AC1654}.
\bcomment{Publisher: IOP Publishing}.
2022-03-01
\end{barticle}
\endbibitem

\bibitem{eich_elm_2017}
\begin{barticle}
\bauthor{\bsnm{Eich}, \binits{T.}},
\bauthor{\bsnm{Sieglin}, \binits{B.}},
\bauthor{\bsnm{Thornton}, \binits{A.J.}},
\bauthor{\bsnm{Faitsch}, \binits{M.}},
\bauthor{\bsnm{Kirk}, \binits{A.}},
\bauthor{\bsnm{Herrmann}, \binits{A.}},
\bauthor{\bsnm{Suttrop}, \binits{W.}}:
\batitle{{ELM} divertor peak energy fluence scaling to {ITER} with data from {JET}, {MAST} and {ASDEX} upgrade}.
\bjtitle{Nuclear Materials and Energy}
\bvolume{12},
\bfpage{84}--\blpage{90}
(\byear{2017}).
\doiurl{10.1016/j.nme.2017.04.014}.
2024-02-26
\end{barticle}
\endbibitem

\bibitem{ViezzerE.2018}
\begin{barticle}
\bauthor{\bsnm{{Viezzer E.}}}:
\batitle{Access and sustainment of naturally {ELM}-free and small-{ELM} regimes}.
\bjtitle{Nuclear Fusion}
(\byear{2018}).
\doiurl{10.1088/1741-4326/aac222}.
2018-09-27
\end{barticle}
\endbibitem

\bibitem{Whytea2010}
\begin{botherref}
\oauthor{\bsnm{Whyte}, \binits{D.G.}},
\oauthor{\bsnm{Hubbard}, \binits{A.E.}},
\oauthor{\bsnm{Hughes}, \binits{J.W.}},
\oauthor{\bsnm{Lipschultz}, \binits{B.}},
\oauthor{\bsnm{Rice}, \binits{J.E.}},
\oauthor{\bsnm{Marmar}, \binits{E.S.}},
\oauthor{\bsnm{Greenwald}, \binits{M.}},
\oauthor{\bsnm{Cziegler}, \binits{I.}},
\oauthor{\bsnm{Dominguez}, \binits{A.}},
\oauthor{\bsnm{Golfinopoulos}, \binits{T.}},
\oauthor{\bsnm{Howard}, \binits{N.}},
\oauthor{\bsnm{Lin}, \binits{L.}},
\oauthor{\bsnm{McDermottb}, \binits{R.M.}},
\oauthor{\bsnm{Porkolab}, \binits{M.}},
\oauthor{\bsnm{Reinke}, \binits{M.L.}}, et al.:
I-mode: {An} {H}-mode energy confinement regime with {L}-mode particle transport in {Alcator} {C}-{Mod}.
Nuclear Fusion
\textbf{50}(10)
(2010).
\doiurl{10.1088/0029-5515/50/10/105005}
\end{botherref}
\endbibitem

\bibitem{bielajew_edge_2022}
\begin{barticle}
\bauthor{\bsnm{Bielajew}, \binits{R.}},
\bauthor{\bsnm{Conway}, \binits{G.D.}},
\bauthor{\bsnm{Griener}, \binits{M.}},
\bauthor{\bsnm{Happel}, \binits{T.}},
\bauthor{\bsnm{Hfler}, \binits{K.}},
\bauthor{\bsnm{Howard}, \binits{N.T.}},
\bauthor{\bsnm{Hubbard}, \binits{A.E.}},
\bauthor{\bsnm{McCarthy}, \binits{W.}},
\bauthor{\bsnm{Cabrera}, \binits{P.A.M.}},
\bauthor{\bsnm{Nishizawa}, \binits{T.}},
\bauthor{\bsnm{Rodriguez-Fernandez}, \binits{P.}},
\bauthor{\bsnm{Silvagni}, \binits{D.}},
\bauthor{\bsnm{Vanovac}, \binits{B.}},
\bauthor{\bsnm{Wendler}, \binits{D.}},
\bauthor{\bsnm{Yoo}, \binits{C.}}, \betal:
\batitle{Edge turbulence measurements in {L}-mode and {I}-mode at {ASDEX} {Upgrade}}.
\bjtitle{Physics of Plasmas}
\bvolume{29}(\bissue{5}),
\bfpage{052504}
(\byear{2022}).
\doiurl{10.1063/5.0088062}.
\bcomment{Publisher: AIP Publishing LLCAIP Publishing}.
2022-05-09
\end{barticle}
\endbibitem

\bibitem{bielajew_edge_2023}
\begin{barticle}
\bauthor{\bsnm{Bielajew}, \binits{R.}},
\bauthor{\bsnm{Plank}, \binits{U.}},
\bauthor{\bsnm{Conway}, \binits{G.D.}},
\bauthor{\bsnm{Hubbard}, \binits{A.E.}},
\bauthor{\bsnm{Rodriguez-Fernandez}, \binits{P.}},
\bauthor{\bsnm{Vanovac}, \binits{B.}},
\bauthor{\bsnm{Yoo}, \binits{C.}},
\bauthor{\bsnm{White}, \binits{A.E.}},
\bauthor{\bsnm{Team}, \binits{t.A.U.}}:
\batitle{Edge radiated temperature fluctuations across confinement regime transitions in favorable and unfavorable drift configurations at {ASDEX} {Upgrade}}.
\bjtitle{Nuclear Fusion}
\bvolume{63}(\bissue{12}),
\bfpage{126022}
(\byear{2023}).
\doiurl{10.1088/1741-4326/acfcc9}.
\bcomment{Publisher: IOP Publishing}.
2023-10-07
\end{barticle}
\endbibitem

\bibitem{Snyder2009}
\begin{botherref}
\oauthor{\bsnm{Snyder}, \binits{P.B.}},
\oauthor{\bsnm{Groebner}, \binits{R.J.}},
\oauthor{\bsnm{Leonard}, \binits{A.W.}},
\oauthor{\bsnm{Osborne}, \binits{T.H.}},
\oauthor{\bsnm{Wilson}, \binits{H.R.}}:
Development and validation of a predictive model for the pedestal height.
Physics of Plasmas
\textbf{16}(5)
(2009).
\doiurl{10.1063/1.3122146}
\end{botherref}
\endbibitem

\bibitem{Hughes2020a}
\begin{barticle}
\bauthor{\bsnm{Hughes}, \binits{J.W.}},
\bauthor{\bsnm{Howard}, \binits{N.T.}},
\bauthor{\bsnm{Rodriguez-Fernandez}, \binits{P.}},
\bauthor{\bsnm{Creely}, \binits{A.J.}},
\bauthor{\bsnm{Kuang}, \binits{A.Q.}},
\bauthor{\bsnm{Snyder}, \binits{P.B.}},
\bauthor{\bsnm{Wilks}, \binits{T.M.}},
\bauthor{\bsnm{Sweeney}, \binits{R.}},
\bauthor{\bsnm{Greenwald}, \binits{M.}}:
\batitle{Projections of {H}-mode access and edge pedestal in the {SPARC} tokamak}.
\bjtitle{Journal of Plasma Physics}
\bvolume{86}(\bissue{5}),
\bfpage{865860504}
(\byear{2020}).
\doiurl{10.1017/S0022377820001300}.
\bcomment{Publisher: Cambridge University Press}.
2021-07-12
\end{barticle}
\endbibitem

\bibitem{Rodriguez-Fernandez2020a}
\begin{barticle}
\bauthor{\bsnm{Rodriguez-Fernandez}, \binits{P.}},
\bauthor{\bsnm{Howard}, \binits{N.T.}},
\bauthor{\bsnm{Greenwald}, \binits{M.J.}},
\bauthor{\bsnm{Creely}, \binits{A.J.}},
\bauthor{\bsnm{Hughes}, \binits{J.W.}},
\bauthor{\bsnm{Wright}, \binits{J.C.}},
\bauthor{\bsnm{Holland}, \binits{C.}},
\bauthor{\bsnm{Lin}, \binits{Y.}},
\bauthor{\bsnm{Sciortino}, \binits{F.}},
\bauthor{\bsnm{Team}, \binits{t.S.}}:
\batitle{Predictions of core plasma performance for the {SPARC} tokamak}.
\bjtitle{Journal of Plasma Physics}
\bvolume{86}(\bissue{5}),
\bfpage{865860503}
(\byear{2020}).
\doiurl{10.1017/S0022377820001075}.
\bcomment{Publisher: Cambridge University Press}.
2021-07-12
\end{barticle}
\endbibitem

\bibitem{battaglia_aps23}
\begin{botherref}
\oauthor{\bsnm{Battaglia}, \binits{D.J.}},
\oauthor{\bsnm{Body}, \binits{T.A.}},
\oauthor{\bsnm{Brookman}, \binits{M.W.}},
\oauthor{\bsnm{Creely}, \binits{A.J.}},
\oauthor{\bsnm{Eich}, \binits{T.}},
\oauthor{\bsnm{Hasse}, \binits{C.}},
\oauthor{\bsnm{Myers}, \binits{C.E.}},
\oauthor{\bsnm{Reinke}, \binits{M.L.}},
\oauthor{\bsnm{Scott}, \binits{S.}},
\oauthor{\bsnm{Sweeney}, \binits{R.M.}},
\oauthor{\bsnm{Howard}, \binits{N.T.}},
\oauthor{\bsnm{Hubbard}, \binits{A.E.}},
\oauthor{\bsnm{Hughes}, \binits{J.W.}},
\oauthor{\bsnm{Perks}, \binits{C.J.}},
\oauthor{\bsnm{Rea}, \binits{C.}},
\oauthor{\bsnm{Rodriguez-Fernandez}, \binits{P.}},
\oauthor{\bsnm{Rice}, \binits{J.E.}},
\oauthor{\bsnm{Saltzman}, \binits{A.}},
\oauthor{\bsnm{Tinguely}, \binits{A.}},
\oauthor{\bsnm{Wilks}, \binits{T.M.}},
\oauthor{\bsnm{Wigram}, \binits{M.}},
\oauthor{\bsnm{Nelson}, \binits{A.O.}},
\oauthor{\bsnm{Paz-Soldan}, \binits{C.A.}},
\oauthor{\bsnm{Logan}, \binits{N.C.}}:
Q $>$ 1 operation space in the first SPARC campaign.
Talk presented at the 65th Annual Meeting of the APS Division of Plasma Physics, abstract NO05.00002.
Denver (CO), November 1
(2023)
\end{botherref}
\endbibitem

\bibitem{body_aps23}
\begin{botherref}
\oauthor{\bsnm{Body}, \binits{T.A.}},
\oauthor{\bsnm{Kuang}, \binits{A.Q.}},
\oauthor{\bsnm{Reinke}, \binits{M.L.}},
\oauthor{\bsnm{Eich}, \binits{T.}},
\oauthor{\bsnm{Looby}, \binits{T.}},
\oauthor{\bsnm{Brookman}, \binits{M.W.}},
\oauthor{\bsnm{Hasse}, \binits{C.}},
\oauthor{\bsnm{Rodriguez-Fernandez}, \binits{P.}},
\oauthor{\bsnm{Howard}, \binits{N.T.}},
\oauthor{\bsnm{Wigram}, \binits{M.}},
\oauthor{\bsnm{Ballinger}, \binits{S.B.}},
\oauthor{\bsnm{Corsaro}, \binits{L.}},
\oauthor{\bsnm{Nelson}, \binits{O.A.}}:
Integrating edge exhaust in SPARC L-mode scenario development.
Talk presented at the 65th Annual Meeting of the APS Division of Plasma Physics, abstract NO05.00003.
Denver (CO), November 1
(2023)
\end{botherref}
\endbibitem

\bibitem{tinguely_aps23}
\begin{botherref}
\oauthor{\bsnm{Tinguely}, \binits{R.A.}},
\oauthor{\bsnm{Reinke}, \binits{M.L.}},
\oauthor{\bsnm{Paz-Soldan}, \binits{C.A.}},
\oauthor{\bsnm{Ball}, \binits{J.L.}},
\oauthor{\bsnm{Granetz}, \binits{R.S.}},
\oauthor{\bsnm{Howard}, \binits{N.T.}},
\oauthor{\bsnm{Mackie}, \binits{S.}},
\oauthor{\bsnm{Panontin}, \binits{E.}},
\oauthor{\bsnm{Perks}, \binits{C.J.}},
\oauthor{\bsnm{Rice}, \binits{J.E.}},
\oauthor{\bsnm{Rodriguez-Fernandez}, \binits{P.}},
\oauthor{\bsnm{Wang}, \binits{X.}},
\oauthor{\bsnm{Battaglia}, \binits{D.J.}},
\oauthor{\bsnm{Creely}, \binits{A.J.}},
\oauthor{\bsnm{Gocht}, \binits{R.}},
\oauthor{\bsnm{Holmes}, \binits{I.}},
\oauthor{\bsnm{Myers}, \binits{C.E.}},
\oauthor{\bsnm{Raj}, \binits{P.}},
\oauthor{\bsnm{Vezinet}, \binits{D.}},
\oauthor{\bsnm{Lachmann}, \binits{A.}},
\oauthor{\bsnm{Leuthold}, \binits{N.}},
\oauthor{\bsnm{Stewart}, \binits{I.G.}}:
Measuring fusion gain Q $>$ 1 in SPARC.
Talk presented at the 65th Annual Meeting of the APS Division of Plasma Physics, abstract NO05.00006.
Denver (CO), November 1
(2023)
\end{botherref}
\endbibitem

\bibitem{rodriguezfernandez_aps23}
\begin{botherref}
\oauthor{\bsnm{Rodriguez-Fernandez}, \binits{P.}},
\oauthor{\bsnm{Howard}, \binits{N.T.}},
\oauthor{\bsnm{Saltzman}, \binits{A.}},
\oauthor{\bsnm{Shoji}, \binits{L.}},
\oauthor{\bsnm{Hughes}, \binits{J.W.}},
\oauthor{\bsnm{Battaglia}, \binits{D.J.}},
\oauthor{\bsnm{Body}, \binits{T.A.}},
\oauthor{\bsnm{Creely}, \binits{A.J.}},
\oauthor{\bsnm{Candy}, \binits{J.}}:
On the direct use of core nonlinear gyrokinetic profile predictions for the planning of burning plasma experiments.
Talk presented at the 65th Annual Meeting of the APS Division of Plasma Physics, abstract GI01.00001.
Denver (CO), October 31
(2023)
\end{botherref}
\endbibitem

\bibitem{Sweeney2020a}
\begin{botherref}
\oauthor{\bsnm{Sweeney}, \binits{R.}},
\oauthor{\bsnm{Creely}, \binits{A.J.}},
\oauthor{\bsnm{Doody}, \binits{J.}},
\oauthor{\bsnm{Flp}, \binits{T.}},
\oauthor{\bsnm{Garnier}, \binits{D.T.}},
\oauthor{\bsnm{Granetz}, \binits{R.}},
\oauthor{\bsnm{Greenwald}, \binits{M.}},
\oauthor{\bsnm{Hesslow}, \binits{L.}},
\oauthor{\bsnm{Irby}, \binits{J.}},
\oauthor{\bsnm{Izzo}, \binits{V.A.}},
\oauthor{\bsnm{Haye}, \binits{R.J.L.}},
\oauthor{\bsnm{Logan}, \binits{N.C.}},
\oauthor{\bsnm{Montes}, \binits{K.}},
\oauthor{\bsnm{Paz-Soldan}, \binits{C.}},
\oauthor{\bsnm{Rea}, \binits{C.}}, et al.:
{MHD} stability and disruptions in the {SPARC} tokamak.
Journal of Plasma Physics
\textbf{86}(5)
(2020).
\doiurl{10.1017/S0022377820001129}.
Publisher: Cambridge University Press.
2021-07-12
\end{botherref}
\endbibitem

\bibitem{mailloux_overview_2022}
\begin{barticle}
\bauthor{\bsnm{Mailloux}, \binits{J.}},
\bauthor{\bsnm{Abid}, \binits{N.}},
\bauthor{\bsnm{Abraham}, \binits{K.}},
\bauthor{\bsnm{Abreu}, \binits{P.}},
\bauthor{\bsnm{Adabonyan}, \binits{O.}},
\bauthor{\bsnm{Adrich}, \binits{P.}},
\bauthor{\bsnm{Afanasev}, \binits{V.}},
\bauthor{\bsnm{Afzal}, \binits{M.}},
\bauthor{\bsnm{Ahlgren}, \binits{T.}},
\bauthor{\bsnm{Aho-Mantila}, \binits{L.}},
\bauthor{\bsnm{Aiba}, \binits{N.}},
\bauthor{\bsnm{Airila}, \binits{M.}},
\bauthor{\bsnm{Akhtar}, \binits{M.}},
\bauthor{\bsnm{Albanese}, \binits{R.}},
\bauthor{\bsnm{Alderson-Martin}, \binits{M.}}, \betal:
\batitle{Overview of {JET} results for optimising {ITER} operation}.
\bjtitle{Nuclear Fusion}
\bvolume{62}(\bissue{4}),
\bfpage{042026}
(\byear{2022}).
\doiurl{10.1088/1741-4326/ac47b4}.
2023-02-10
\end{barticle}
\endbibitem

\bibitem{french_construction_1983}
\begin{barticle}
\bauthor{\bsnm{French}, \binits{J.W.}},
\bauthor{\bsnm{Fedor}, \binits{B.J.}},
\bauthor{\bsnm{Shaw}, \binits{L.E.}},
\bauthor{\bsnm{Sabado}, \binits{M.M.}}:
\batitle{Construction of the {Tokamak} {Fusion} {Test} {Reactor}}.
\bjtitle{Nuclear Technology - Fusion}
\bvolume{4}(\bissue{2P2}),
\bfpage{326}--\blpage{335}
(\byear{1983}).
\doiurl{10.13182/FST83-A22887}.
\bcomment{Publisher: Taylor \& Francis \_eprint: https://doi.org/10.13182/FST83-A22887}.
2024-01-20
\end{barticle}
\endbibitem

\bibitem{kishimoto_advanced_2005}
\begin{barticle}
\bauthor{\bsnm{Kishimoto}, \binits{H.}},
\bauthor{\bsnm{Ishida}, \binits{S.}},
\bauthor{\bsnm{Kikuchi}, \binits{M.}},
\bauthor{\bsnm{Ninomiya}, \binits{H.}}:
\batitle{Advanced tokamak research on {JT}-60}.
\bjtitle{Nuclear Fusion}
\bvolume{45}(\bissue{8}),
\bfpage{986}
(\byear{2005}).
\doiurl{10.1088/0029-5515/45/8/026}.
2024-01-20
\end{barticle}
\endbibitem

\bibitem{rodriguez-fernandez_nonlinear_2022}
\begin{barticle}
\bauthor{\bsnm{Rodriguez-Fernandez}, \binits{P.}},
\bauthor{\bsnm{Howard}, \binits{N.T.}},
\bauthor{\bsnm{Candy}, \binits{J.}}:
\batitle{Nonlinear gyrokinetic predictions of {SPARC} burning plasma profiles enabled by surrogate modeling}.
\bjtitle{Nuclear Fusion}
\bvolume{62}(\bissue{7}),
\bfpage{076036}
(\byear{2022}).
\doiurl{10.1088/1741-4326/AC64B2}.
\bcomment{Publisher: IOP Publishing}.
2022-05-16
\end{barticle}
\endbibitem

\bibitem{Houlberg1982}
\begin{botherref}
\oauthor{\bsnm{Houlberg}, \binits{W.A.}}:
Fusion {Reactor} {Plasma} {Performance} {Modeling} - {POPCON} analysis.
Report
(1982)
\end{botherref}
\endbibitem

\bibitem{Yushmanov1999}
\begin{barticle}
\bauthor{\bsnm{Yushmanov}, \binits{P.N.}},
\bauthor{\bsnm{Takizuka}, \binits{T.}},
\bauthor{\bsnm{Riedel}, \binits{K.S.}},
\bauthor{\bsnm{Kardaun}, \binits{O.}},
\bauthor{\bsnm{Cordey}, \binits{J.G.}},
\bauthor{\bsnm{Kaye}, \binits{S.M.}},
\bauthor{\bsnm{Post}, \binits{D.E.}}:
\batitle{Scalings for tokamak energy confinement}.
\bjtitle{Nuclear Fusion}
\bvolume{30}(\bissue{10}),
\bfpage{1990}
(\byear{1990})
\end{barticle}
\endbibitem

\bibitem{Rice2020}
\begin{botherref}
\oauthor{\bsnm{Rice}, \binits{J.E.}},
\oauthor{\bsnm{Citrin}, \binits{J.}},
\oauthor{\bsnm{Cao}, \binits{N.M.}},
\oauthor{\bsnm{Diamond}, \binits{P.H.}},
\oauthor{\bsnm{Greenwald}, \binits{M.}},
\oauthor{\bsnm{Grierson}, \binits{B.A.}}:
Understanding {LOC}/{SOC} phenomenology in tokamaks.
Nuclear Fusion
\textbf{60}(10)
(2020).
\doiurl{10.1088/1741-4326/abac4b}
\end{botherref}
\endbibitem

\bibitem{mitim}
\begin{botherref}
\oauthor{\bsnm{Rodriguez-Fernandez}, \binits{P.}}, et al.:
MITIM: a toolbox for modeling tasks in plasma physics and fusion energy.
\url{https://github.com/pabloprf/MITIM-fusion}.
Version 1.1
(2024).
\url{https://mitim-fusion.readthedocs.io/en/latest/}
\end{botherref}
\endbibitem

\bibitem{balandat_botorch_2020}
\begin{botherref}
\oauthor{\bsnm{Balandat}, \binits{M.}},
\oauthor{\bsnm{Karrer}, \binits{B.}},
\oauthor{\bsnm{Jiang}, \binits{D.R.}},
\oauthor{\bsnm{Daulton}, \binits{S.}},
\oauthor{\bsnm{Letham}, \binits{B.}},
\oauthor{\bsnm{Wilson}, \binits{A.G.}},
\oauthor{\bsnm{Bakshy}, \binits{E.}}:
{BoTorch}: {A} {Framework} for {Efficient} {Monte}-{Carlo} {Bayesian} {Optimization}.
arXiv.
arXiv:1910.06403 [cs, math, stat]
(2020).
\doiurl{10.48550/arXiv.1910.06403}.
\url{http://arxiv.org/abs/1910.06403}
2023-12-19
\end{botherref}
\endbibitem

\bibitem{Candy2016}
\begin{barticle}
\bauthor{\bsnm{Candy}, \binits{J.}},
\bauthor{\bsnm{Belli}, \binits{E.A.}},
\bauthor{\bsnm{Bravenec}, \binits{R.V.}}:
\batitle{A high-accuracy {Eulerian} gyrokinetic solver for collisional plasmas}.
\bjtitle{Journal of Computational Physics}
\bvolume{324},
\bfpage{73}--\blpage{93}
(\byear{2016}).
\doiurl{10.1016/j.jcp.2016.07.039}.
\bcomment{Publisher: Elsevier Inc.}
\end{barticle}
\endbibitem

\bibitem{howard_prediction_2024}
\begin{botherref}
\oauthor{\bsnm{Howard}, \binits{N.T.}},
\oauthor{\bsnm{Rodriguez-Fernandez}, \binits{P.}},
\oauthor{\bsnm{Holland}, \binits{C.}},
\oauthor{\bsnm{Candy}, \binits{J.}}:
Prediction of {Performance} and {Turbulence} in {ITER} {Burning} {Plasmas} via {Nonlinear} {Gyrokinetic} {Profile} {Prediction}.
arXiv.
arXiv:2404.17040 [physics]
(2024).
\doiurl{10.48550/arXiv.2404.17040}.
\url{http://arxiv.org/abs/2404.17040}
2024-05-03
\end{botherref}
\endbibitem

\bibitem{howard_simultaneous_2024}
\begin{barticle}
\bauthor{\bsnm{Howard}, \binits{N.T.}},
\bauthor{\bsnm{Rodriguez-Fernandez}, \binits{P.}},
\bauthor{\bsnm{Holland}, \binits{C.}},
\bauthor{\bsnm{Odstrcil}, \binits{T.}},
\bauthor{\bsnm{Grierson}, \binits{B.}},
\bauthor{\bsnm{Sciortino}, \binits{F.}},
\bauthor{\bsnm{McKee}, \binits{G.}},
\bauthor{\bsnm{Yan}, \binits{Z.}},
\bauthor{\bsnm{Wang}, \binits{G.}},
\bauthor{\bsnm{Rhodes}, \binits{T.L.}},
\bauthor{\bsnm{White}, \binits{A.E.}},
\bauthor{\bsnm{Candy}, \binits{J.}},
\bauthor{\bsnm{Chrystal}, \binits{C.}}:
\batitle{Simultaneous reproduction of experimental profiles, fluxes, transport coefficients, and turbulence characteristics via nonlinear gyrokinetic profile predictions in a {DIII}-{D} {ITER} similar shape plasma}.
\bjtitle{Physics of Plasmas}
\bvolume{31}(\bissue{3}),
\bfpage{032501}
(\byear{2024}).
\doiurl{10.1063/5.0175792}.
2024-03-01
\end{barticle}
\endbibitem

\bibitem{holland_aps23}
\begin{botherref}
\oauthor{\bsnm{Holland}, \binits{C.G.}}:
Characterization of Predicted Confinement and Transport in an ARC-class Tokamak Power Plant.
Talk presented at the 65th Annual Meeting of the APS Division of Plasma Physics, abstract GI01.00002.
Denver (CO), October 31
(2023)
\end{botherref}
\endbibitem

\bibitem{rodriguezfernandez_eps23}
\begin{botherref}
\oauthor{\bsnm{Rodriguez-Fernandez}, \binits{P.}},
\oauthor{\bsnm{Howard}, \binits{N.T.}},
\oauthor{\bsnm{Delabie}, \binits{E.}},
\oauthor{\bsnm{Lomanowski}, \binits{B.}},
\oauthor{\bsnm{Saltzman}, \binits{A.}},
\oauthor{\bsnm{Kantamneni}, \binits{S.}},
\oauthor{\bsnm{Candy}, \binits{J.}},
\oauthor{\bsnm{Holland}, \binits{C.}},
\oauthor{\bsnm{Nave}, \binits{M.F.F.}},
\oauthor{\bsnm{Biewer}, \binits{T.M.}},
\oauthor{\bsnm{Garcia}, \binits{J.}},
\oauthor{\bsnm{Lennholm}, \binits{M.}},
\oauthor{\bsnm{White}, \binits{A.E.}},
\oauthor{\bsnm{{JET Contributors}}}:
Prediction of core kinetic profiles and burning plasma performance with high-fidelity gyrokinetic simulations in tokamaks.
Talk presented at the 49th European Conference on Plasma Physics.
Bordeaux, France, July 3-7
(2023)
\end{botherref}
\endbibitem

\bibitem{rodriguezfernandez_2024_jet}
\begin{botherref}
\oauthor{\bsnm{Rodriguez-Fernandez1}, \binits{P.}},
\oauthor{\bsnm{Delabie}, \binits{E.}},
\oauthor{\bsnm{Howard}, \binits{N.T.}},
\oauthor{\bsnm{Lomanowski}, \binits{B.}},
\oauthor{\bsnm{Nave}, \binits{M.F.F.}},
\oauthor{\bsnm{Biewer}, \binits{T.M.}},
\oauthor{\bsnm{White}, \binits{A.E.}},
\oauthor{\bsnm{Garcia}, \binits{J.}},
\oauthor{\bsnm{Lennholm}, \binits{M.}},
\oauthor{\bsnm{Contributors}, \binits{J.}}:
Study of the isotope effect in jet ohmic plasmas via direct nonlinear gyrokinetic profile prediction.
Plasma Physics and Controlled Fusion (submitted)
(2024)
\end{botherref}
\endbibitem

\bibitem{holland_examination_2021}
\begin{barticle}
\bauthor{\bsnm{Holland}, \binits{C.}},
\bauthor{\bsnm{Luce}, \binits{T.C.}},
\bauthor{\bsnm{Grierson}, \binits{B.A.}},
\bauthor{\bsnm{Smith}, \binits{S.P.}},
\bauthor{\bsnm{Marinoni}, \binits{A.}},
\bauthor{\bsnm{Burrell}, \binits{K.H.}},
\bauthor{\bsnm{Petty}, \binits{C.C.}},
\bauthor{\bsnm{Bass}, \binits{E.M.}}:
\batitle{Examination of stiff ion temperature gradient mode physics in simulations of {DIII}-{D} {H}-mode transport}.
\bjtitle{Nuclear Fusion}
\bvolume{61}(\bissue{6}),
\bfpage{066033}
(\byear{2021}).
\doiurl{10.1088/1741-4326/abf951}.
\bcomment{Publisher: IOP Publishing}.
2024-02-08
\end{barticle}
\endbibitem

\bibitem{candy_multiscale-optimized_2019}
\begin{barticle}
\bauthor{\bsnm{Candy}, \binits{J.}},
\bauthor{\bsnm{Sfiligoi}, \binits{I.}},
\bauthor{\bsnm{Belli}, \binits{E.}},
\bauthor{\bsnm{Hallatschek}, \binits{K.}},
\bauthor{\bsnm{Holland}, \binits{C.}},
\bauthor{\bsnm{Howard}, \binits{N.}},
\bauthor{\bsnm{DAzevedo}, \binits{E.}}:
\batitle{Multiscale-optimized plasma turbulence simulation on petascale architectures}.
\bjtitle{Computers \& Fluids}
\bvolume{188},
\bfpage{125}--\blpage{135}
(\byear{2019}).
\doiurl{10.1016/j.compfluid.2019.04.016}.
2024-02-08
\end{barticle}
\endbibitem

\bibitem{Staebler2005}
\begin{barticle}
\bauthor{\bsnm{Staebler}, \binits{G.M.}},
\bauthor{\bsnm{Kinsey}, \binits{J.E.}},
\bauthor{\bsnm{Waltz}, \binits{R.E.}}:
\batitle{Gyro-{Landau} fluid equations for trapped and passing particles}.
\bjtitle{Physics of Plasmas}
\bvolume{12}(\bissue{10}),
\bfpage{1}--\blpage{24}
(\byear{2005}).
\doiurl{10.1063/1.2044587}
\end{barticle}
\endbibitem

\bibitem{Staebler2007}
\begin{barticle}
\bauthor{\bsnm{Staebler}, \binits{G.M.}},
\bauthor{\bsnm{Kinsey}, \binits{J.E.}},
\bauthor{\bsnm{Waltz}, \binits{R.E.}}:
\batitle{A theory-based transport model with comprehensive physics}.
\bjtitle{Physics of Plasmas}
\bvolume{14}(\bissue{5}),
\bfpage{055909}
(\byear{2007}).
\doiurl{10.1063/1.2436852}
\end{barticle}
\endbibitem

\bibitem{Belli2008}
\begin{barticle}
\bauthor{\bsnm{Belli}, \binits{E.A.}},
\bauthor{\bsnm{Candy}, \binits{J.}}:
\batitle{Kinetic calculation of neoclassical transport including self-consistent electron and impurity dynamics}.
\bjtitle{Plasma Physics and Controlled Fusion}
\bvolume{50}(\bissue{9}),
\bfpage{095010}
(\byear{2008}).
\doiurl{10.1088/0741-3335/50/9/095010}
\end{barticle}
\endbibitem

\bibitem{Breslau2018}
\begin{barticle}
\bauthor{\bsnm{Breslau}, \binits{J.}},
\bauthor{\bsnm{Gorelenkova}, \binits{M.}},
\bauthor{\bsnm{Poli}, \binits{F.}},
\bauthor{\bsnm{Sachdev}, \binits{J.}},
\bauthor{\bsnm{Pankin}, \binits{A.}},
\bauthor{\bsnm{Perumpilly}, \binits{G.}},
\bauthor{\bsnm{Yuan}, \binits{X.}},
\bauthor{\bsnm{Glant}, \binits{L.}}:
\batitle{{TRANSP}}.
\bjtitle{Computer software. USDOE Office of Science (SC), Fusion Energy Sciences (FES)}
(\byear{2018}).
\doiurl{10.11578/DC.20180627.4}
\end{barticle}
\endbibitem

\bibitem{Brambilla1999}
\begin{botherref}
\oauthor{\bsnm{Brambilla}, \binits{M.}}:
Numerical simulation of ion cyclotron waves in tokamak plasmas.
Plasma Phys. Control. Fusion
\textbf{41}(1)
(1999)
\end{botherref}
\endbibitem

\bibitem{Hammett1986}
\begin{botherref}
\oauthor{\bsnm{Hammett}, \binits{G.}}:
Fast ion studies of ion cyclotron heating in the {PLT} tokamak: {Phd} thesises.
PhD thesis,
Princeton University
(1986)
\end{botherref}
\endbibitem

\bibitem{angioni_dependence_2023}
\begin{barticle}
\bauthor{\bsnm{Angioni}, \binits{C.}},
\bauthor{\bsnm{Bonanomi}, \binits{N.}},
\bauthor{\bsnm{Fable}, \binits{E.}},
\bauthor{\bsnm{Schneider}, \binits{P.A.}},
\bauthor{\bsnm{Tardini}, \binits{G.}},
\bauthor{\bsnm{Luda}, \binits{T.}},
\bauthor{\bsnm{Staebler}, \binits{G.M.}},
\bauthor{\bsnm{Team}, \binits{t.A.U.}}:
\batitle{The dependence of tokamak {L}-mode confinement on magnetic field and plasma size, from a magnetic field scan experiment at {ASDEX} {Upgrade} to full-radius integrated modelling and fusion reactor predictions}.
\bjtitle{Nuclear Fusion}
\bvolume{63}(\bissue{5}),
\bfpage{056005}
(\byear{2023}).
\doiurl{10.1088/1741-4326/acc193}.
\bcomment{Publisher: IOP Publishing}.
2023-03-31
\end{barticle}
\endbibitem

\bibitem{Angioni2022}
\begin{barticle}
\bauthor{\bsnm{Angioni}, \binits{C.}},
\bauthor{\bsnm{Gamot}, \binits{T.}},
\bauthor{\bsnm{Tardini}, \binits{G.}},
\bauthor{\bsnm{Fable}, \binits{E.}},
\bauthor{\bsnm{Luda}, \binits{T.}},
\bauthor{\bsnm{Bonanomi}, \binits{N.}},
\bauthor{\bsnm{Kiefer}, \binits{C.}},
\bauthor{\bsnm{Staebler}, \binits{G.}},
\bauthor{\bsnm{ASDEX Upgrade~Team}, \binits{T.}},
\bauthor{\bsnm{Mst}, \binits{E.}}:
\batitle{Confinement properties of {L}-mode plasmas in {ASDEX} {Upgrade} and full-radius predictions of the {TGLF} transport model}.
\bjtitle{Nuclear Fusion}
\bvolume{62},
\bfpage{066015}
(\byear{2022}).
\doiurl{10.1088/1741-4326/AC592B}.
\bcomment{Publisher: IOP Publishing}.
2022-03-11
\end{barticle}
\endbibitem

\bibitem{hughes_thomson_2003}
\begin{barticle}
\bauthor{\bsnm{Hughes}, \binits{J.W.}},
\bauthor{\bsnm{Mossessian}, \binits{D.}},
\bauthor{\bsnm{Zhurovich}, \binits{K.}},
\bauthor{\bsnm{DeMaria}, \binits{M.}},
\bauthor{\bsnm{Jensen}, \binits{K.}},
\bauthor{\bsnm{Hubbard}, \binits{A.}}:
\batitle{Thomson scattering upgrades on {Alcator} {C}-{Mod}}.
\bjtitle{Review of Scientific Instruments}
\bvolume{74}(\bissue{3}),
\bfpage{1667}--\blpage{1670}
(\byear{2003}).
\doiurl{10.1063/1.1532764}.
2023-10-18
\end{barticle}
\endbibitem

\bibitem{Kotschenreuther2019}
\begin{barticle}
\bauthor{\bsnm{Kotschenreuther}, \binits{M.}},
\bauthor{\bsnm{Liu}, \binits{X.}},
\bauthor{\bsnm{Hatch}, \binits{D.R.}},
\bauthor{\bsnm{Mahajan}, \binits{S.}},
\bauthor{\bsnm{Zheng}, \binits{L.}},
\bauthor{\bsnm{Diallo}, \binits{A.}},
\bauthor{\bsnm{Groebner}, \binits{R.}},
\bauthor{\bsnm{Hillesheim}, \binits{J.C.}},
\bauthor{\bsnm{Maggi}, \binits{C.F.}},
\bauthor{\bsnm{Giroud}, \binits{C.}},
\bauthor{\bsnm{Koechl}, \binits{F.}},
\bauthor{\bsnm{Parail}, \binits{V.}},
\bauthor{\bsnm{Saarelma}, \binits{S.}},
\bauthor{\bsnm{Solano}, \binits{E.}},
\bauthor{\bsnm{Chankin}, \binits{A.}}:
\batitle{Gyrokinetic analysis and simulation of pedestals to identify the culprits for energy losses using fingerprints}.
\bjtitle{Nuclear Fusion}
\bvolume{59}(\bissue{9}),
\bfpage{096001}
(\byear{2019}).
\doiurl{10.1088/1741-4326/AB1FA2}.
\bcomment{Publisher: IOP Publishing}.
2022-02-02
\end{barticle}
\endbibitem

\bibitem{Arbon2020}
\begin{barticle}
\bauthor{\bsnm{Arbon}, \binits{R.}},
\bauthor{\bsnm{Candy}, \binits{J.}},
\bauthor{\bsnm{Belli}, \binits{E.A.}}:
\batitle{Rapidly-convergent flux-surface shape parameterization}.
\bjtitle{Plasma Physics and Controlled Fusion}
\bvolume{63}(\bissue{1}),
\bfpage{012001}
(\byear{2020}).
\doiurl{10.1088/1361-6587/abc63b}
\end{barticle}
\endbibitem

\bibitem{sugama_nonlinear_1998}
\begin{barticle}
\bauthor{\bsnm{Sugama}, \binits{H.}},
\bauthor{\bsnm{Horton}, \binits{W.}}:
\batitle{Nonlinear electromagnetic gyrokinetic equation for plasmas with large mean flows}.
\bjtitle{Physics of Plasmas}
\bvolume{5}(\bissue{7}),
\bfpage{2560}--\blpage{2573}
(\byear{1998}).
\doiurl{10.1063/1.872941}.
2023-11-30
\end{barticle}
\endbibitem

\bibitem{Sauter2014}
\begin{botherref}
\oauthor{\bsnm{Sauter}, \binits{O.}},
\oauthor{\bsnm{Brunner}, \binits{S.}},
\oauthor{\bsnm{Kim}, \binits{D.}},
\oauthor{\bsnm{Merlo}, \binits{G.}},
\oauthor{\bsnm{Behn}, \binits{R.}},
\oauthor{\bsnm{Camenen}, \binits{Y.}},
\oauthor{\bsnm{Coda}, \binits{S.}},
\oauthor{\bsnm{Duval}, \binits{B.P.}},
\oauthor{\bsnm{Federspiel}, \binits{L.}},
\oauthor{\bsnm{Goodman}, \binits{T.P.}},
\oauthor{\bsnm{Karpushov}, \binits{A.}},
\oauthor{\bsnm{Merle}, \binits{A.}},
\oauthor{\bsnm{Team}, \binits{T.}}:
On the non-stiffness of edge transport in {L}-mode tokamak plasmas.
Physics of Plasmas
\textbf{21}(5)
(2014).
\doiurl{10.1063/1.4876612}
\end{botherref}
\endbibitem

\bibitem{kaye_iter_1997}
\begin{barticle}
\bauthor{\bsnm{Kaye}, \binits{S.M.}},
\bauthor{\bsnm{Greenwald}, \binits{M.}},
\bauthor{\bsnm{Stroth}, \binits{U.}},
\bauthor{\bsnm{Kardaun}, \binits{O.}},
\bauthor{\bsnm{Kus}, \binits{A.}},
\bauthor{\bsnm{Schissel}, \binits{D.}},
\bauthor{\bsnm{DeBoo}, \binits{J.}},
\bauthor{\bsnm{Bracco}, \binits{G.}},
\bauthor{\bsnm{Thomsen}, \binits{K.}},
\bauthor{\bsnm{Cordey}, \binits{J.G.}},
\bauthor{\bsnm{Miura}, \binits{Y.}},
\bauthor{\bsnm{Matsuda}, \binits{T.}},
\bauthor{\bsnm{Tamai}, \binits{H.}},
\bauthor{\bsnm{Takizuda}, \binits{T.}},
\bauthor{\bsnm{Hirayama}, \binits{T.}}, \betal:
\batitle{{ITER} {L} mode confinement database}.
\bjtitle{Nuclear Fusion}
\bvolume{37}(\bissue{9}),
\bfpage{1303}--\blpage{1328}
(\byear{1997}).
\doiurl{10.1088/0029-5515/37/9/I10}.
\bcomment{Publisher: IOP Publishing}.
2022-09-12
\end{barticle}
\endbibitem

\bibitem{holland_development_2023}
\begin{barticle}
\bauthor{\bsnm{Holland}, \binits{C.}},
\bauthor{\bsnm{Bass}, \binits{E.M.}},
\bauthor{\bsnm{Orlov}, \binits{D.M.}},
\bauthor{\bsnm{McClenaghan}, \binits{J.}},
\bauthor{\bsnm{Lyons}, \binits{B.C.}},
\bauthor{\bsnm{Grierson}, \binits{B.A.}},
\bauthor{\bsnm{Jian}, \binits{X.}},
\bauthor{\bsnm{Howard}, \binits{N.T.}},
\bauthor{\bsnm{Rodriguez-Fernandez}, \binits{P.}}:
\batitle{Development of compact tokamak fusion reactor use cases to inform future transport studies}.
\bjtitle{Journal of Plasma Physics}
\bvolume{89}(\bissue{4}),
\bfpage{905890418}
(\byear{2023}).
\doiurl{10.1017/S0022377823000843}.
2023-08-30
\end{barticle}
\endbibitem

\bibitem{kato_energy_2024}
\begin{botherref}
\oauthor{\bsnm{Kato}, \binits{T.}},
\oauthor{\bsnm{Sugama}, \binits{H.}},
\oauthor{\bsnm{Watanabe}, \binits{T.-H.}},
\oauthor{\bsnm{Nunami}, \binits{M.}}:
Energy exchange between electrons and ions in ion temperature gradient turbulence.
arXiv.
arXiv:2402.12748 [physics]
(2024).
\doiurl{10.48550/arXiv.2402.12748}.
\url{http://arxiv.org/abs/2402.12748}
2024-05-01
\end{botherref}
\endbibitem

\bibitem{Angioni2007}
\begin{barticle}
\bauthor{\bsnm{Angioni}, \binits{C.}},
\bauthor{\bsnm{Weisen}, \binits{H.}},
\bauthor{\bsnm{Kardaun}, \binits{O.J.W.F.}},
\bauthor{\bsnm{Maslov}, \binits{M.}},
\bauthor{\bsnm{Zabolotsky}, \binits{A.}},
\bauthor{\bsnm{Fuchs}, \binits{C.}},
\bauthor{\bsnm{Garzotti}, \binits{L.}},
\bauthor{\bsnm{Giroud}, \binits{C.}},
\bauthor{\bsnm{Kurzan}, \binits{B.}},
\bauthor{\bsnm{Mantica}, \binits{P.}},
\bauthor{\bsnm{Peeters}, \binits{A.G.}},
\bauthor{\bsnm{Stober}, \binits{J.}}:
\batitle{Scaling of density peaking in {H}-mode plasmas based on a combined database of {AUG} and {JET} observations}.
\bjtitle{Nuclear Fusion}
\bvolume{47}(\bissue{9}),
\bfpage{1326}--\blpage{1335}
(\byear{2007}).
\doiurl{10.1088/0029-5515/47/9/033}
\end{barticle}
\endbibitem

\bibitem{Greenwald2007a}
\begin{barticle}
\bauthor{\bsnm{Greenwald}, \binits{M.}},
\bauthor{\bsnm{Angioni}, \binits{C.}},
\bauthor{\bsnm{Hughes}, \binits{J.W.}},
\bauthor{\bsnm{Terry}, \binits{J.}},
\bauthor{\bsnm{Weisen}, \binits{H.}}:
\batitle{Density profile peaking in low collisionality {H}-modes: comparison of {Alcator} {C}-{Mod} data to {ASDEX} {Upgrade}/{JET} scalings}.
\bjtitle{Nuclear Fusion}
\bvolume{47}(\bissue{9}),
\bfpage{26}
(\byear{2007}).
\doiurl{10.1088/0029-5515/47/9/L03}.
\bcomment{Publisher: IOP Publishing}.
2021-10-19
\end{barticle}
\endbibitem

\bibitem{mckee_impurity-induced_2000}
\begin{barticle}
\bauthor{\bsnm{McKee}, \binits{G.}},
\bauthor{\bsnm{Burrell}, \binits{K.}},
\bauthor{\bsnm{Fonck}, \binits{R.}},
\bauthor{\bsnm{Jackson}, \binits{G.}},
\bauthor{\bsnm{Murakami}, \binits{M.}},
\bauthor{\bsnm{Staebler}, \binits{G.}},
\bauthor{\bsnm{Thomas}, \binits{D.}},
\bauthor{\bsnm{West}, \binits{P.}}:
\batitle{Impurity-{Induced} {Suppression} of {Core} {Turbulence} and {Transport} in the {DIII}-{D} {Tokamak}}.
\bjtitle{Physical Review Letters}
\bvolume{84}(\bissue{9}),
\bfpage{1922}--\blpage{1925}
(\byear{2000}).
\doiurl{10.1103/PhysRevLett.84.1922}.
\bcomment{Publisher: American Physical Society}.
2024-05-02
\end{barticle}
\endbibitem

\bibitem{McKee2000}
\begin{barticle}
\bauthor{\bsnm{McKee}, \binits{G.R.}},
\bauthor{\bsnm{Murakami}, \binits{M.}},
\bauthor{\bsnm{Boedo}, \binits{J.A.}},
\bauthor{\bsnm{Brooks}, \binits{N.H.}},
\bauthor{\bsnm{Burrell}, \binits{K.H.}},
\bauthor{\bsnm{Ernst}, \binits{D.R.}},
\bauthor{\bsnm{Fonck}, \binits{R.J.}},
\bauthor{\bsnm{Jackson}, \binits{G.L.}},
\bauthor{\bsnm{Jakubowski}, \binits{M.}},
\bauthor{\bsnm{La~Haye}, \binits{R.J.}},
\bauthor{\bsnm{Messiaen}, \binits{A.M.}},
\bauthor{\bsnm{Ongena}, \binits{J.}},
\bauthor{\bsnm{Rettig}, \binits{C.L.}},
\bauthor{\bsnm{Rice}, \binits{B.W.}},
\bauthor{\bsnm{Rost}, \binits{C.}}, \betal:
\batitle{Impurity-induced turbulence suppression and reduced transport in the {DIII}-{D} tokamak}.
\bjtitle{Physics of Plasmas}
\bvolume{7}(\bissue{5 II}),
\bfpage{1870}--\blpage{1877}
(\byear{2000}).
\doiurl{10.1063/1.874010}
\end{barticle}
\endbibitem

\bibitem{ennever_effects_2015}
\begin{barticle}
\bauthor{\bsnm{Ennever}, \binits{P.}},
\bauthor{\bsnm{Porkolab}, \binits{M.}},
\bauthor{\bsnm{Candy}, \binits{J.}},
\bauthor{\bsnm{Staebler}, \binits{G.}},
\bauthor{\bsnm{Reinke}, \binits{M.L.}},
\bauthor{\bsnm{Rice}, \binits{J.E.}},
\bauthor{\bsnm{Rost}, \binits{J.C.}},
\bauthor{\bsnm{Ernst}, \binits{D.}},
\bauthor{\bsnm{Fiore}, \binits{C.}},
\bauthor{\bsnm{Hughes}, \binits{J.}},
\bauthor{\bsnm{Terry}, \binits{J.}},
\bauthor{\bsnm{{Alcator C-Mod Team}}}:
\batitle{The effects of dilution on turbulence and transport in {C}-{Mod} ohmic plasmas and comparisons with gyrokinetic simulations}.
\bjtitle{Physics of Plasmas}
\bvolume{22}(\bissue{7}),
\bfpage{072507}
(\byear{2015}).
\doiurl{10.1063/1.4926518}.
2022-11-28
\end{barticle}
\endbibitem

\bibitem{hatch_gyrokinetic_2017}
\begin{barticle}
\bauthor{\bsnm{Hatch}, \binits{D.R.}},
\bauthor{\bsnm{Kotschenreuther}, \binits{M.}},
\bauthor{\bsnm{Mahajan}, \binits{S.}},
\bauthor{\bsnm{Valanju}, \binits{P.}},
\bauthor{\bsnm{Liu}, \binits{X.}}:
\batitle{A gyrokinetic perspective on the {JET}-{ILW} pedestal}.
\bjtitle{Nuclear Fusion}
\bvolume{57}(\bissue{3}),
\bfpage{036020}
(\byear{2017}).
\doiurl{10.1088/1741-4326/aa51e1}.
\bcomment{Publisher: IOP Publishing}.
2024-05-02
\end{barticle}
\endbibitem

\bibitem{li_impurity_2019}
\begin{barticle}
\bauthor{\bsnm{Li}, \binits{J.}},
\bauthor{\bsnm{Wang}, \binits{Z.X.}},
\bauthor{\bsnm{Dong}, \binits{J.Q.}},
\bauthor{\bsnm{Han}, \binits{M.K.}},
\bauthor{\bsnm{Shen}, \binits{Y.}},
\bauthor{\bsnm{Xiao}, \binits{Y.}},
\bauthor{\bsnm{Du}, \binits{H.R.}}:
\batitle{Impurity effects on ion temperature gradient driven multiple modes in transport barriers}.
\bjtitle{Nuclear Fusion}
\bvolume{59}(\bissue{7}),
\bfpage{076013}
(\byear{2019}).
\doiurl{10.1088/1741-4326/ab0ee2}.
\bcomment{Publisher: IOP Publishing}.
2024-05-02
\end{barticle}
\endbibitem

\bibitem{Staebler2017}
\begin{botherref}
\oauthor{\bsnm{Staebler}, \binits{G.M.M.}},
\oauthor{\bsnm{Howard}, \binits{N.T.}},
\oauthor{\bsnm{Candy}, \binits{J.}},
\oauthor{\bsnm{Holland}, \binits{C.}}:
A model of the saturation of coupled electron and ion scale gyrokinetic turbulence.
Nuclear Fusion
\textbf{57}(6)
(2017).
\doiurl{10.1088/1741-4326/aa6bee}.
Publisher: IOP Publishing
\end{botherref}
\endbibitem

\bibitem{Staebler2020}
\begin{barticle}
\bauthor{\bsnm{Staebler}, \binits{G.M.}},
\bauthor{\bsnm{Candy}, \binits{J.}},
\bauthor{\bsnm{Belli}, \binits{E.A.}},
\bauthor{\bsnm{Kinsey}, \binits{J.E.}},
\bauthor{\bsnm{Bonanomi}, \binits{N.}},
\bauthor{\bsnm{Patel}, \binits{B.}}:
\batitle{Geometry dependence of the fluctuation intensity in gyrokinetic turbulence}.
\bjtitle{Plasma Physics and Controlled Fusion}
\bvolume{63}(\bissue{1}),
\bfpage{015013}
(\byear{2020}).
\doiurl{10.1088/1361-6587/abc861}
\end{barticle}
\endbibitem

\bibitem{Staebler2021}
\begin{barticle}
\bauthor{\bsnm{Staebler}, \binits{G.M.}},
\bauthor{\bsnm{Belli}, \binits{E.A.}},
\bauthor{\bsnm{Candy}, \binits{J.}},
\bauthor{\bsnm{Kinsey}, \binits{J.E.}},
\bauthor{\bsnm{Dudding}, \binits{H.}},
\bauthor{\bsnm{Patel}, \binits{B.}}:
\batitle{Verification of a quasi-linear model for gyrokinetic turbulent transport}.
\bjtitle{Nuclear Fusion}
\bvolume{61},
\bfpage{116007}
(\byear{2021}).
\doiurl{10.1088/1741-4326/ac243a}.
2021-12-25
\end{barticle}
\endbibitem

\bibitem{dudding_new_2022}
\begin{botherref}
\oauthor{\bsnm{Dudding}, \binits{H.}}:
A new quasilinear saturation rule for tokamak turbulence.
PhD thesis,
University of York
(2022)
\end{botherref}
\endbibitem

\bibitem{Creely2019}
\begin{barticle}
\bauthor{\bsnm{Creely}, \binits{A.J.}},
\bauthor{\bsnm{Rodriguez-Fernandez}, \binits{P.}},
\bauthor{\bsnm{Conway}, \binits{G.D.}},
\bauthor{\bsnm{Freethy}, \binits{S.J.}},
\bauthor{\bsnm{Howard}, \binits{N.T.}},
\bauthor{\bsnm{White}, \binits{A.E.}}:
\batitle{Criteria for the importance of multi-scale interactions in turbulent transport simulations}.
\bjtitle{Plasma Phys. Control. Fusion}
\bvolume{61},
\bfpage{085022}
(\byear{2019}).
\bcomment{Publisher: IOP Publishing}
\end{barticle}
\endbibitem

\bibitem{rodriguezfernandez_aps21}
\begin{botherref}
\oauthor{\bsnm{Rodriguez-Fernandez}, \binits{P.}},
\oauthor{\bsnm{Howard}, \binits{N.T.}},
\oauthor{\bsnm{Creely}, \binits{A.J.}},
\oauthor{\bsnm{Spector}, \binits{B.F.}},
\oauthor{\bsnm{Greenwald}, \binits{M.J.}},
\oauthor{\bsnm{Hughes}, \binits{J.W.}}:
Physics exploration of scenarios towards breakeven and burning plasmas in the SPARC tokamak.
Talk presented at the 63rd Annual Meeting of the APS Division of Plasma Physics, abstract JO07.00003.
Pittsburgh (PA), November 9
(2021)
\end{botherref}
\endbibitem

\bibitem{hughes_fec23}
\begin{bchapter}
\bauthor{\bsnm{Hughes}, \binits{J.}},
\bauthor{\bsnm{Creely}, \binits{A.}},
\bauthor{\bsnm{Hubbard}, \binits{A.}},
\bauthor{\bsnm{Battaglia}, \binits{D.}},
\bauthor{\bsnm{Howard}, \binits{N.}},
\bauthor{\bsnm{Rodriguez-Fernandez}, \binits{P.}},
\bauthor{\bsnm{Wilks}, \binits{T.}}:
\bctitle{Access to edge transport barriers and projections of pedestal performance in the sparc tokamak}.
In: \bbtitle{Proceedings of the 29th Fusion Energy Conference (FEC 2023)},
\bconflocation{London, UK}
(\byear{2023}).
\bcomment{International Atomic Energy Agency (IAEA)}.
\burl{https://conferences.iaea.org/event/316/contributions/28663/}
\end{bchapter}
\endbibitem

\bibitem{Ryter2014}
\begin{botherref}
\oauthor{\bsnm{Ryter}, \binits{F.}},
\oauthor{\bsnm{Barrera~Orte}, \binits{L.}},
\oauthor{\bsnm{Kurzan}, \binits{B.}},
\oauthor{\bsnm{McDermott}, \binits{R.M.}},
\oauthor{\bsnm{Tardini}, \binits{G.}},
\oauthor{\bsnm{Viezzer}, \binits{E.}},
\oauthor{\bsnm{Bernert}, \binits{M.}},
\oauthor{\bsnm{Fischer}, \binits{R.}}:
Experimental evidence for the key role of the ion heat channel in the physics of the {L}-{H} transition.
Nuclear Fusion
\textbf{54}(8)
(2014).
\doiurl{10.1088/0029-5515/54/8/083003}
\end{botherref}
\endbibitem

\bibitem{ryter_i-mode_2016}
\begin{barticle}
\bauthor{\bsnm{Ryter}, \binits{F.}},
\bauthor{\bsnm{Fischer}, \binits{R.}},
\bauthor{\bsnm{Fuchs}, \binits{J.C.}},
\bauthor{\bsnm{Happel}, \binits{T.}},
\bauthor{\bsnm{McDermott}, \binits{R.M.}},
\bauthor{\bsnm{Viezzer}, \binits{E.}},
\bauthor{\bsnm{Wolfrum}, \binits{E.}},
\bauthor{\bsnm{Orte}, \binits{L.B.}},
\bauthor{\bsnm{Bernert}, \binits{M.}},
\bauthor{\bsnm{Burckhart}, \binits{A.}},
\bauthor{\bsnm{Graa}, \binits{S.d.}},
\bauthor{\bsnm{Kurzan}, \binits{B.}},
\bauthor{\bsnm{McCarthy}, \binits{P.}},
\bauthor{\bsnm{Ptterich}, \binits{T.}},
\bauthor{\bsnm{Suttrop}, \binits{W.}}, \betal:
\batitle{I-mode studies at {ASDEX} {Upgrade}: {L}-{I} and {I}-{H} transitions, pedestal and confinement properties}.
\bjtitle{Nuclear Fusion}
\bvolume{57}(\bissue{1}),
\bfpage{016004}
(\byear{2016}).
\doiurl{10.1088/0029-5515/57/1/016004}.
\bcomment{Publisher: IOP Publishing}.
2024-02-06
\end{barticle}
\endbibitem

\bibitem{siena_predictions_2023}
\begin{barticle}
\bauthor{\bsnm{Siena}, \binits{A.D.}},
\bauthor{\bsnm{Rodriguez-Fernandez}, \binits{P.}},
\bauthor{\bsnm{Howard}, \binits{N.T.}},
\bauthor{\bsnm{Navarro}, \binits{A.B.}},
\bauthor{\bsnm{Bilato}, \binits{R.}},
\bauthor{\bsnm{Grler}, \binits{T.}},
\bauthor{\bsnm{Poli}, \binits{E.}},
\bauthor{\bsnm{Merlo}, \binits{G.}},
\bauthor{\bsnm{Wright}, \binits{J.}},
\bauthor{\bsnm{Greenwald}, \binits{M.}},
\bauthor{\bsnm{Jenko}, \binits{F.}}:
\batitle{Predictions of improved confinement in {SPARC} via energetic particle turbulence stabilization}.
\bjtitle{Nuclear Fusion}
\bvolume{63}(\bissue{3}),
\bfpage{036003}
(\byear{2023}).
\doiurl{10.1088/1741-4326/acb1c7}.
\bcomment{Publisher: IOP Publishing}.
2023-02-01
\end{barticle}
\endbibitem

\bibitem{Howard2021}
\begin{barticle}
\bauthor{\bsnm{Howard}, \binits{N.T.}},
\bauthor{\bsnm{Rodriguez-Fernandez}, \binits{P.} \bsuffix{P.}},
\bauthor{\bsnm{Holland}, \binits{C.}},
\bauthor{\bsnm{Rice}, \binits{J.E.}},
\bauthor{\bsnm{Greenwald}, \binits{M.}},
\bauthor{\bsnm{Candy}, \binits{J.}},
\bauthor{\bsnm{Sciortino}, \binits{F.}}:
\batitle{Gyrokinetic simulation of turbulence and transport in the {SPARC} tokamak}.
\bjtitle{Physics of Plasmas}
\bvolume{28},
\bfpage{072502}
(\byear{2021}).
\doiurl{10.1063/5.0047789}.
\bcomment{Publisher: AIP Publishing LLC}
\end{barticle}
\endbibitem

\bibitem{Sorbom2015}
\begin{barticle}
\bauthor{\bsnm{Sorbom}, \binits{B.N.}},
\bauthor{\bsnm{Ball}, \binits{J.}},
\bauthor{\bsnm{Palmer}, \binits{T.R.}},
\bauthor{\bsnm{Mangiarotti}, \binits{F.J.}},
\bauthor{\bsnm{Sierchio}, \binits{J.M.}},
\bauthor{\bsnm{Bonoli}, \binits{P.}},
\bauthor{\bsnm{Kasten}, \binits{C.}},
\bauthor{\bsnm{Sutherland}, \binits{D.A.}},
\bauthor{\bsnm{Barnard}, \binits{H.S.}},
\bauthor{\bsnm{Haakonsen}, \binits{C.B.}},
\bauthor{\bsnm{Goh}, \binits{J.}},
\bauthor{\bsnm{Sung}, \binits{C.}},
\bauthor{\bsnm{Whyte}, \binits{D.G.}}:
\batitle{{ARC}: {A} compact, high-field, fusion nuclear science facility and demonstration power plant with demountable magnets}.
\bjtitle{Fusion Engineering and Design}
\bvolume{100},
\bfpage{378}--\blpage{405}
(\byear{2015}).
\doiurl{10.1016/j.fusengdes.2015.07.008}.
\bcomment{Publisher: Elsevier B.V.}
\end{barticle}
\endbibitem

\bibitem{frank_radiative_2022}
\begin{barticle}
\bauthor{\bsnm{Frank}, \binits{S.J.}},
\bauthor{\bsnm{Perks}, \binits{C.J.}},
\bauthor{\bsnm{Nelson}, \binits{A.O.}},
\bauthor{\bsnm{Qian}, \binits{T.}},
\bauthor{\bsnm{Jin}, \binits{S.}},
\bauthor{\bsnm{Cavallaro}, \binits{A.}},
\bauthor{\bsnm{Rutkowski}, \binits{A.}},
\bauthor{\bsnm{Reiman}, \binits{A.}},
\bauthor{\bsnm{Freidberg}, \binits{J.P.}},
\bauthor{\bsnm{Rodriguez-Fernandez}, \binits{P.}},
\bauthor{\bsnm{Whyte}, \binits{D.}}:
\batitle{Radiative pulsed {L}-mode operation in {ARC}-class reactors}.
\bjtitle{Nuclear Fusion}
\bvolume{62}(\bissue{12}),
\bfpage{126036}
(\byear{2022}).
\doiurl{10.1088/1741-4326/ac95ac}.
\bcomment{Publisher: IOP Publishing}.
2022-11-07
\end{barticle}
\endbibitem

\bibitem{wilson_aps23}
\begin{botherref}
\oauthor{\bsnm{Wilson}, \binits{H.S.}},
\oauthor{\bsnm{Arnold}, \binits{D.A.}},
\oauthor{\bsnm{Boyes}, \binits{W.}},
\oauthor{\bsnm{Chandra}, \binits{R.N.}},
\oauthor{\bsnm{Choudhury}, \binits{H.P.}},
\oauthor{\bsnm{DaSilva}, \binits{N.J.}},
\oauthor{\bsnm{Hansen}, \binits{C.J.}},
\oauthor{\bsnm{Liu}, \binits{Y.}},
\oauthor{\bsnm{Lunia}, \binits{P.}},
\oauthor{\bsnm{Nelson}, \binits{A.O.}},
\oauthor{\bsnm{Paz-Soldan}, \binits{C.A.}},
\oauthor{\bsnm{Pharr}, \binits{M.C.}},
\oauthor{\bsnm{Tobin}, \binits{M.}},
\oauthor{\bsnm{Ball}, \binits{J.L.}},
\oauthor{\bsnm{Bielajew}, \binits{R.}},
\oauthor{\bsnm{Benjamin}, \binits{S.R.}},
\oauthor{\bsnm{Calvo-Carrera}, \binits{M.}},
\oauthor{\bsnm{Chang}, \binits{C.W.}},
\oauthor{\bsnm{Corsaro}, \binits{L.}},
\oauthor{\bsnm{Cummings}, \binits{C.}},
\oauthor{\bsnm{Devitre}, \binits{A.}},
\oauthor{\bsnm{Diab}, \binits{R.}},
\oauthor{\bsnm{Ferry}, \binits{S.E.}},
\oauthor{\bsnm{Frank}, \binits{S.}},
\oauthor{\bsnm{Jerkins}, \binits{J.}},
\oauthor{\bsnm{Johnson}, \binits{J.}},
\oauthor{\bsnm{Mackie}, \binits{S.P.}},
\oauthor{\bsnm{Mandell}, \binits{N.R.}},
\oauthor{\bsnm{Maris}, \binits{A.}},
\oauthor{\bsnm{Miller}, \binits{M.A.}},
\oauthor{\bsnm{Mouratidis}, \binits{T.}},
\oauthor{\bsnm{Murphy}, \binits{D.}},
\oauthor{\bsnm{Peterson}, \binits{E.E.}},
\oauthor{\bsnm{Rutherford}, \binits{G.}},
\oauthor{\bsnm{Saltzman}, \binits{A.}},
\oauthor{\bsnm{Segantin}, \binits{S.}},
\oauthor{\bparticle{van~de} \bsnm{Lindt}, \binits{J.G.}},
\oauthor{\bsnm{Velberg}, \binits{A.}},
\oauthor{\bsnm{Wang}, \binits{A.M.}},
\oauthor{\bsnm{Wigram}, \binits{M.}},
\oauthor{\bsnm{Witham}, \binits{J.}},
\oauthor{\bsnm{Whyte}, \binits{D.}}:
Integrated Design of a Pulsed, Modular, Negative Triangularity Pilot Plant Concept.
Poster presented at the 65th Annual Meeting of the APS Division of Plasma Physics, abstract BP11.00131.
Denver (CO), November 1
(2023)
\end{botherref}
\endbibitem

\bibitem{White2019}
\begin{barticle}
\bauthor{\bsnm{White}, \binits{A.E.}}:
\batitle{Validation of nonlinear gyrokinetic transport models using turbulence measurements}.
\bjtitle{Journal of Plasma Physics}
\bvolume{85}(\bissue{1}),
\bfpage{925850102}
(\byear{2019}).
\doiurl{10.1017/S0022377818001253}
\end{barticle}
\endbibitem

\bibitem{fable_high-confinement_2021}
\begin{barticle}
\bauthor{\bsnm{Fable}, \binits{E.}},
\bauthor{\bsnm{Kallenbach}, \binits{A.}},
\bauthor{\bsnm{McDermott}, \binits{R.M.}},
\bauthor{\bsnm{Bernert}, \binits{M.}},
\bauthor{\bsnm{Angioni}, \binits{C.}},
\bauthor{\bsnm{Team}, \binits{t.A.U.}}:
\batitle{High-confinement radiative {L}-modes in {ASDEX} {Upgrade}}.
\bjtitle{Nuclear Fusion}
\bvolume{62}(\bissue{2}),
\bfpage{024001}
(\byear{2021}).
\doiurl{10.1088/1741-4326/ac3e81}.
\bcomment{Publisher: IOP Publishing}.
2023-02-21
\end{barticle}
\endbibitem

\end{thebibliography}

\end{document}